\documentclass[onecolumn]{aastex63}
\usepackage{xcolor}

\usepackage{natbib} 
\usepackage{amsmath}
\usepackage{enumitem}
\shorttitle{TNOs in the first four years of \des}
\shortauthors{Bernardinelli et al.}

\defcitealias{Bernstein2000}{BK}

\newcommand{\E}{\mathrm{E}}
\newcommand{\ra}{\mathrm{RA}}
\newcommand{\dec}{\mathrm{Dec}}
\newcommand{\des}{\textit{DES}}

\newcommand{\AU}{\,\mathrm{AU}}

\begin{document}

\title{Trans-Neptunian objects found in the first four years of the Dark Energy Survey}

\author[0000-0003-0743-9422]{Pedro H. Bernardinelli}
\affiliation{Department of Physics and Astronomy, University of Pennsylvania, Philadelphia, PA 19104, USA}
\email{pedrobe@sas.upenn.edu}

\author[0000-0002-8613-8259]{Gary M. Bernstein}
\affiliation{Department of Physics and Astronomy, University of Pennsylvania, Philadelphia, PA 19104, USA}

\author[0000-0003-2764-7093]{Masao Sako}
\affiliation{Department of Physics and Astronomy, University of Pennsylvania, Philadelphia, PA 19104, USA}

\author[0000-0001-5677-5188]{Tongtian Liu}
\affiliation{Department of Physics and Astronomy, University of Pennsylvania, Philadelphia, PA 19104, USA}

\author[0000-0002-8737-742X]{William R. Saunders}
\affiliation{Department of Physics and Astronomy, University of Pennsylvania, Philadelphia, PA 19104, USA}
\affiliation{Department of Astronomy, Boston University, Boston, MA 02215, USA}

\author[0000-0001-7721-6457]{Tali Khain} 
\affiliation{Department of Physics, University of Michigan, Ann Arbor, MI 48109, USA}

\author[0000-0001-7737-6784]{Hsing Wen Lin} 
\affiliation{Department of Physics, University of Michigan, Ann Arbor, MI 48109, USA}

\author[0000-0001-6942-2736]{David W. Gerdes} 
\affiliation{Department of Astronomy, University of Michigan, Ann Arbor, MI 48109, USA}
\affiliation{Department of Physics, University of Michigan, Ann Arbor, MI 48109, USA}

\author[0000-0001-5201-8374]{Dillon Brout}
\affiliation{Department of Physics and Astronomy, University of Pennsylvania, Philadelphia, PA 19104, USA}

\author[0000-0002-8167-1767]{Fred C. Adams}
\affiliation{Department of Astronomy, University of Michigan, Ann Arbor, MI 48109, USA}
\affiliation{Department of Physics, University of Michigan, Ann Arbor, MI 48109, USA}

\author{Matthew Belyakov}
\affiliation{Department of Physics and Astronomy, University of Pennsylvania, Philadelphia, PA 19104, USA}

\author{Aditya Inada Somasundaram}
\affiliation{Department of Physics and Astronomy, University of Pennsylvania, Philadelphia, PA 19104, USA}

\author{Lakshay Sharma}
\affiliation{Department of Physics and Astronomy, University of Pennsylvania, Philadelphia, PA 19104, USA}

\author{Jennifer Locke}
\affiliation{Department of Physics and Astronomy, University of Pennsylvania, Philadelphia, PA 19104, USA}

\author{Kyle Franson} 
\affiliation{Department of Astronomy, The University of Texas at Austin, Austin, TX 78712, USA}
\affiliation{Department of Physics, University of Michigan, Ann Arbor, MI 48109, USA}

\author[0000-0002-7733-4522]{Juliette C. Becker} 
\affiliation{Department of Astronomy, University of Michigan, Ann Arbor, MI 48109, USA}

\author{Kevin Napier} 
\affiliation{Department of Physics, University of Michigan, Ann Arbor, MI 48109, USA}

\author[0000-0002-2486-1118]{Larissa Markwardt} 
\affiliation{Department of Astronomy, University of Michigan, Ann Arbor, MI 48109, USA}

\author[0000-0002-0609-3987]{James Annis}
\affiliation{Fermi National Accelerator Laboratory, P.O. Box 500, Batavia, IL 60510, USA}

\author{T.~M.~C.~Abbott}
\affiliation{Cerro Tololo Inter-American Observatory, National Optical-Infrared Astronomy Observatory, Casilla 603, La Serena, Chile}
\author{S.~Avila}
\affiliation{Instituto de Fisica Teorica UAM/CSIC, Universidad Autonoma de Madrid, 28049 Madrid, Spain}
\author{D.~Brooks}
\affiliation{Department of Physics \& Astronomy, University College London, Gower Street, London, WC1E 6BT, UK}
\author{D.~L.~Burke}
\affiliation{Kavli Institute for Particle Astrophysics \& Cosmology, P. O. Box 2450, Stanford University, Stanford, CA 94305, USA}
\affiliation{SLAC National Accelerator Laboratory, Menlo Park, CA 94025, USA}
\author{A.~Carnero~Rosell}
\affiliation{Centro de Investigaciones Energ\'eticas, Medioambientales y Tecnol\'ogicas (CIEMAT), Madrid, Spain}
\affiliation{Laborat\'orio Interinstitucional de e-Astronomia - LIneA, Rua Gal. Jos\'e Cristino 77, Rio de Janeiro, RJ - 20921-400, Brazil}
\author{M.~Carrasco~Kind}
\affiliation{Department of Astronomy, University of Illinois at Urbana-Champaign, 1002 W. Green Street, Urbana, IL 61801, USA}
\affiliation{National Center for Supercomputing Applications, 1205 West Clark St., Urbana, IL 61801, USA}
\author{F.~J.~Castander}
\affiliation{Institut d'Estudis Espacials de Catalunya (IEEC), 08034 Barcelona, Spain}
\affiliation{Institute of Space Sciences (ICE, CSIC),  Campus UAB, Carrer de Can Magrans, s/n,  08193 Barcelona, Spain}
\author{L.~N.~da Costa}
\affiliation{Laborat\'orio Interinstitucional de e-Astronomia - LIneA, Rua Gal. Jos\'e Cristino 77, Rio de Janeiro, RJ - 20921-400, Brazil}
\affiliation{Observat\'orio Nacional, Rua Gal. Jos\'e Cristino 77, Rio de Janeiro, RJ - 20921-400, Brazil}
\author{J.~De~Vicente}
\affiliation{Centro de Investigaciones Energ\'eticas, Medioambientales y Tecnol\'ogicas (CIEMAT), Madrid, Spain}
\author{S.~Desai}
\affiliation{Department of Physics, IIT Hyderabad, Kandi, Telangana 502285, India}
\author{H.~T.~Diehl}
\affiliation{Fermi National Accelerator Laboratory, P. O. Box 500, Batavia, IL 60510, USA}
\author{P.~Doel}
\affiliation{Department of Physics \& Astronomy, University College London, Gower Street, London, WC1E 6BT, UK}
\author{S.~Everett}
\affiliation{Santa Cruz Institute for Particle Physics, Santa Cruz, CA 95064, USA}
\author{B.~Flaugher}
\affiliation{Fermi National Accelerator Laboratory, P. O. Box 500, Batavia, IL 60510, USA}
\author{J.~Garc\'ia-Bellido}
\affiliation{Instituto de Fisica Teorica UAM/CSIC, Universidad Autonoma de Madrid, 28049 Madrid, Spain}
\author{D.~Gruen}
\affiliation{Department of Physics, Stanford University, 382 Via Pueblo Mall, Stanford, CA 94305, USA}
\affiliation{Kavli Institute for Particle Astrophysics \& Cosmology, P. O. Box 2450, Stanford University, Stanford, CA 94305, USA}
\affiliation{SLAC National Accelerator Laboratory, Menlo Park, CA 94025, USA}
\author{R.~A.~Gruendl}
\affiliation{Department of Astronomy, University of Illinois at Urbana-Champaign, 1002 W. Green Street, Urbana, IL 61801, USA}
\affiliation{National Center for Supercomputing Applications, 1205 West Clark St., Urbana, IL 61801, USA}
\author{J.~Gschwend}
\affiliation{Laborat\'orio Interinstitucional de e-Astronomia - LIneA, Rua Gal. Jos\'e Cristino 77, Rio de Janeiro, RJ - 20921-400, Brazil}
\affiliation{Observat\'orio Nacional, Rua Gal. Jos\'e Cristino 77, Rio de Janeiro, RJ - 20921-400, Brazil}
\author{G.~Gutierrez}
\affiliation{Fermi National Accelerator Laboratory, P. O. Box 500, Batavia, IL 60510, USA}
\author{D.~L.~Hollowood}
\affiliation{Santa Cruz Institute for Particle Physics, Santa Cruz, CA 95064, USA}
\author{D.~J.~James}
\affiliation{Center for Astrophysics $\vert$ Harvard \& Smithsonian, 60 Garden Street, Cambridge, MA 02138, USA}
\author{M.~W.~G.~Johnson}
\affiliation{National Center for Supercomputing Applications, 1205 West Clark St., Urbana, IL 61801, USA}
\author{M.~D.~Johnson}
\affiliation{National Center for Supercomputing Applications, 1205 West Clark St., Urbana, IL 61801, USA}
\author{E.~Krause}
\affiliation{Department of Astronomy/Steward Observatory, University of Arizona, 933 North Cherry Avenue, Tucson, AZ 85721-0065, USA}
\author{N.~Kuropatkin}
\affiliation{Fermi National Accelerator Laboratory, P. O. Box 500, Batavia, IL 60510, USA}
\author{M.~A.~G.~Maia}
\affiliation{Laborat\'orio Interinstitucional de e-Astronomia - LIneA, Rua Gal. Jos\'e Cristino 77, Rio de Janeiro, RJ - 20921-400, Brazil}
\affiliation{Observat\'orio Nacional, Rua Gal. Jos\'e Cristino 77, Rio de Janeiro, RJ - 20921-400, Brazil}
\author{M.~March}
\affiliation{Department of Physics and Astronomy, University of Pennsylvania, Philadelphia, PA 19104, USA}
\author{R.~Miquel}
\affiliation{Instituci\'o Catalana de Recerca i Estudis Avan\c{c}ats, E-08010 Barcelona, Spain}
\affiliation{Institut de F\'{\i}sica d'Altes Energies (IFAE), The Barcelona Institute of Science and Technology, Campus UAB, 08193 Bellaterra (Barcelona) Spain}
\author{F.~Paz-Chinch\'{o}n}
\affiliation{Department of Astronomy, University of Illinois at Urbana-Champaign, 1002 W. Green Street, Urbana, IL 61801, USA}
\affiliation{National Center for Supercomputing Applications, 1205 West Clark St., Urbana, IL 61801, USA}
\author{A.~A.~Plazas}
\affiliation{Department of Astrophysical Sciences, Princeton University, Peyton Hall, Princeton, NJ 08544, USA}
\author{A.~K.~Romer}
\affiliation{Department of Physics and Astronomy, Pevensey Building, University of Sussex, Brighton, BN1 9QH, UK}
\author{E.~S.~Rykoff}
\affiliation{Kavli Institute for Particle Astrophysics \& Cosmology, P. O. Box 2450, Stanford University, Stanford, CA 94305, USA}
\affiliation{SLAC National Accelerator Laboratory, Menlo Park, CA 94025, USA}
\author{C.~S{\'a}nchez}
\affiliation{Department of Physics and Astronomy, University of Pennsylvania, Philadelphia, PA 19104, USA}
\author{E.~Sanchez}
\affiliation{Centro de Investigaciones Energ\'eticas, Medioambientales y Tecnol\'ogicas (CIEMAT), Madrid, Spain}
\author{V.~Scarpine}
\affiliation{Fermi National Accelerator Laboratory, P. O. Box 500, Batavia, IL 60510, USA}
\author{S.~Serrano}
\affiliation{Institut d'Estudis Espacials de Catalunya (IEEC), 08034 Barcelona, Spain}
\affiliation{Institute of Space Sciences (ICE, CSIC),  Campus UAB, Carrer de Can Magrans, s/n,  08193 Barcelona, Spain}
\author{I.~Sevilla-Noarbe}
\affiliation{Centro de Investigaciones Energ\'eticas, Medioambientales y Tecnol\'ogicas (CIEMAT), Madrid, Spain}
\author{M.~Smith}
\affiliation{School of Physics and Astronomy, University of Southampton,  Southampton, SO17 1BJ, UK}
\author{F.~Sobreira}
\affiliation{Instituto de F\'isica Gleb Wataghin, Universidade Estadual de Campinas, 13083-859, Campinas, SP, Brazil}
\affiliation{Laborat\'orio Interinstitucional de e-Astronomia - LIneA, Rua Gal. Jos\'e Cristino 77, Rio de Janeiro, RJ - 20921-400, Brazil}
\author{E.~Suchyta}
\affiliation{Computer Science and Mathematics Division, Oak Ridge National Laboratory, Oak Ridge, TN 37831}
\author{M.~E.~C.~Swanson}
\affiliation{National Center for Supercomputing Applications, 1205 West Clark St., Urbana, IL 61801, USA}
\author{G.~Tarle}
\affiliation{Department of Physics, University of Michigan, Ann Arbor, MI 48109, USA}
\author{A.~R.~Walker}
\affiliation{Cerro Tololo Inter-American Observatory, National Optical-Infrared Astronomy Observatory, Casilla 603, La Serena, Chile}
\author{W.~Wester}
\affiliation{Fermi National Accelerator Laboratory, P. O. Box 500, Batavia, IL 60510, USA}
\author{Y.~Zhang}
\affiliation{Fermi National Accelerator Laboratory, P. O. Box 500, Batavia, IL 60510, USA}

\collaboration{1000}{(The \des\ Collaboration)\vspace{0.4in}}

\begin{abstract}
We present a catalog of 316 trans-Neptunian bodies (TNOs) detected from the first four seasons (``Y4'' data) of the \emph{Dark Energy Survey} (\des).  The survey covers a contiguous 5000~deg$^2$ of the southern sky in the $grizY$ optical/NIR filter set, with a typical TNO in this part of the sky being targeted by $25-30$ Y4 exposures. This paper focusses on the methods used to detect these objects from the $\approx60,000$ Y4 exposures, a process made challenging by the absence of the few-hour repeat observations employed by TNO-optimized surveys.
Newly developed techniques include: transient/moving object detection by comparison of single-epoch catalogs to catalogs of ``stacked'' images; quantified astrometric error from atmospheric turbulence; new software for detecting TNO linkages in a temporally sparse transient catalog, and for estimating the rate of spurious linkages; and use of faint stars to determine the detection efficiency vs magnitude in all exposures.  Final validation of the reality of linked orbits uses a new ``sub-threshold confirmation'' test, wherein we demand the object be detectable in a stack of the exposures in which the orbit indicates an object should be present, but was not individually detected.
This catalog contains all validated TNOs which were detected on $\ge 6$ unique nights in the Y4 data, and is complete to $r\lesssim 23.3$~mag with virtually no dependence on orbital properties for bound TNOs at distance $30\,{\rm AU}<d<2500\,{\rm AU}.$  
The catalog includes 245 discoveries by \des, 139 not previously published.
The final \des\ TNO catalog is expected to yield $>0.3$~mag more depth, and arcs of $>4$~years for nearly all detections.
\end{abstract}
\reportnum{DES-2019-0447}
\reportnum{FERMILAB-PUB-19-446-AE}

\section{Introduction} \label{sec:intro}
Trans-Neptunian objects (TNOs) are probes of the dynamical and chemical history of the solar system. These planetesimals are relics of major dynamical events among and beyond the giant planets, with the current observed orbital distribution of the Kuiper Belt being a signature of large scale changes in the positions of the giant planets \citep{Fernandez1984,Malhotra1993,Duncan1995,Hahn2005,Tsiganis2005,Levison2008,Nesvorny2015,Nesvorny2016,Kaib2016}. By constraining the detailed structure of the multiple populations in the distant solar system \citep[see][for a review]{Gladman2008}, one can further probe numerous dynamical processes, such as instabilities in Neptune's orbit \citep{Dawson2012}, interactions between these planetesimals and Neptune \citep{Gomes2003,Morbidelli2008}, the presence of distant planetary-mass perturbers \citep{Trujillo2014,Batygin2016,Volk2017}, the effect of close stellar encounters \citep{Jilkova2015}, the birth cluster of the Solar System \citep{Adams2010,Brasser2015}, and perturbations from Galactic tides \citep{Duncan2008,Bannister2017}.

Searches for TNOs face a trade-off between depth and search area.  As the resources available for TNO searches, quantified by the product of (FOV) $\times$ (telescope area) $\times$ (observing time), have increased due to improved detector technology, the envelope of searches in the depth-area plane has expanded.   Dedicated TNO surveys using large-format CCD cameras on 4- or 8-meter-class telescopes now cover hundreds to thousands of square degrees \citep{Bannister2016, Weryk2016, Bannister2018,Chen2018, 2019AJ....157..139S}. The \textit{Dark Energy Survey} (\des) was allocated 575 nights of time on the 4-meter Blanco Telescope in Cerro Tololo over 6 seasons from 2013--2019, with the primary goals of characterizing the distribution of dark matter and the nature of the Hubble acceleration \citep{des_3x2,des_h0,des_bao,des_sn,PhysRevLett.122.171301}.  The survey strategy is, nominally, to image the same contiguous 5000~deg$^2$ patch of high-Galactic-latitude Southern sky in five optical/NIR bands each year.  While not optimized for TNO discovery, the survey is nonetheless capable of pushing out the depth/area envelope of TNO searches, particularly for high-inclination TNOs.  The Dark Energy Camera \citep[][DECam]{Flaugher2015}, with its 3~deg$^2$ field of view and 520 Mpix science array, is a powerful instrument for survey science.  Among previous large-scale surveys optimized for TNO detection is the \textit{Deep Ecliptic Survey} \citep{Elliot2005}.  We caution the reader that, although the acronym \textit{DES} is the same as our survey and both surveys made use of the Blanco telescope, there is no connection with the work presented herein. The \emph{Dark Energy Survey} wide survey covers $10\times$ more area with $\approx 1$~mag deeper TNO detection threshold than the \textit{Deep Ecliptic Survey}, using $\approx 5$ times more nights of 4-meter time, a consequence of the higher quantum efficiency and larger field of view offered by DECam. Discoveries of individual objects of interest have been reported from the \des\ data 
\citep{Gerdes2016,Gerdes2017,Becker2018,Khain2018,Lin2019} as well as from DECam observations allocated for directed TNO searches \citep{Trujillo2014,Sheppard2016,SheppardTrujillo2016,Sheppard2018}.
Astrometric data from \des\ observations of known TNOs have been incorporated into forecasts of future occultation events by \citet{BandaHuarca}.

Here we provide the first comprehensive inventory of TNOs detected in the \des\ observations, from analysis of the first 4 seasons.  We also describe the improvements that are expected when the full survey data are searched (the final observations were in January 2019).

Surveys designed to detect TNOs almost invariably schedule pairs of observations of a sky region with time intervals that are at least an hour, but less than about a day, apart \citep[\textit{e.g.}][]{Jewitt1993,Allen2001,Trujillo2001,Bernstein2004,Elliot2005,Jones2006,Fraser2010,Schwamb2010,Petit2011,Rabinowitz2012,Bannister2016,Sheppard2016,Alexandersen2016,Weryk2016,Chen2018} [note that \citet{Tombaugh} used a few-day interval].
In this regime, the apparent motion of a TNO is large enough to be readily identified in $\approx1\arcsec$ seeing, but the motion is small enough ($\lesssim1\arcmin$)  that two detections of the same TNO are readily identified as such.  The rate of such nearby pairs that are \emph{not} TNOs is small (\textit{e.g.} a single asteroid near turnaround, or coincident detections of two distinct asteroids).  Most large-scale small-body search algorithms rely on the initial identification of such pairs \citep[\textit{e.g.}][]{Denneau2013,Holman2018}.

The \des\ observing strategy specifically \emph{avoids} repeat observations of the same filter/field combination on the same
night, so that the variations in weather are spread across the survey footprint, and the final survey is more valuable if it is more homogeneous.  There is, therefore, no useful way to distinguish TNOs from asteroids in the catalog of $>20$ million transient unresolved \des\ sources.  In this paper, we define a \emph{transient} to be a source that appears in the sky in a given location on only a single night, thus including both moving objects and (non-repeating) variable sources.  We must, like \citet{Perdelwitz2018}, devise algorithms for linking together those detections corresponding to the same TNO.  The linking process, while harder than for a TNO-optimized survey, is fully feasible with proper use of spatial-temporal tree structures for the transients, \textit{e.g.} as in \cite{KUBICA2007151}.  A high-efficiency and high-purity search of the \des\ data is possible because any given TNO is targeted nominally twice per filter per season.  This highly redundant search, while not maximizing the number of TNO discoveries per observation, does mean that the \des\ cadence automatically yields high-quality orbits and multi-band magnitudes for nearly all detected TNOs.

\begin{figure}[t]
	\centering
	\includegraphics[width=0.8\textwidth]{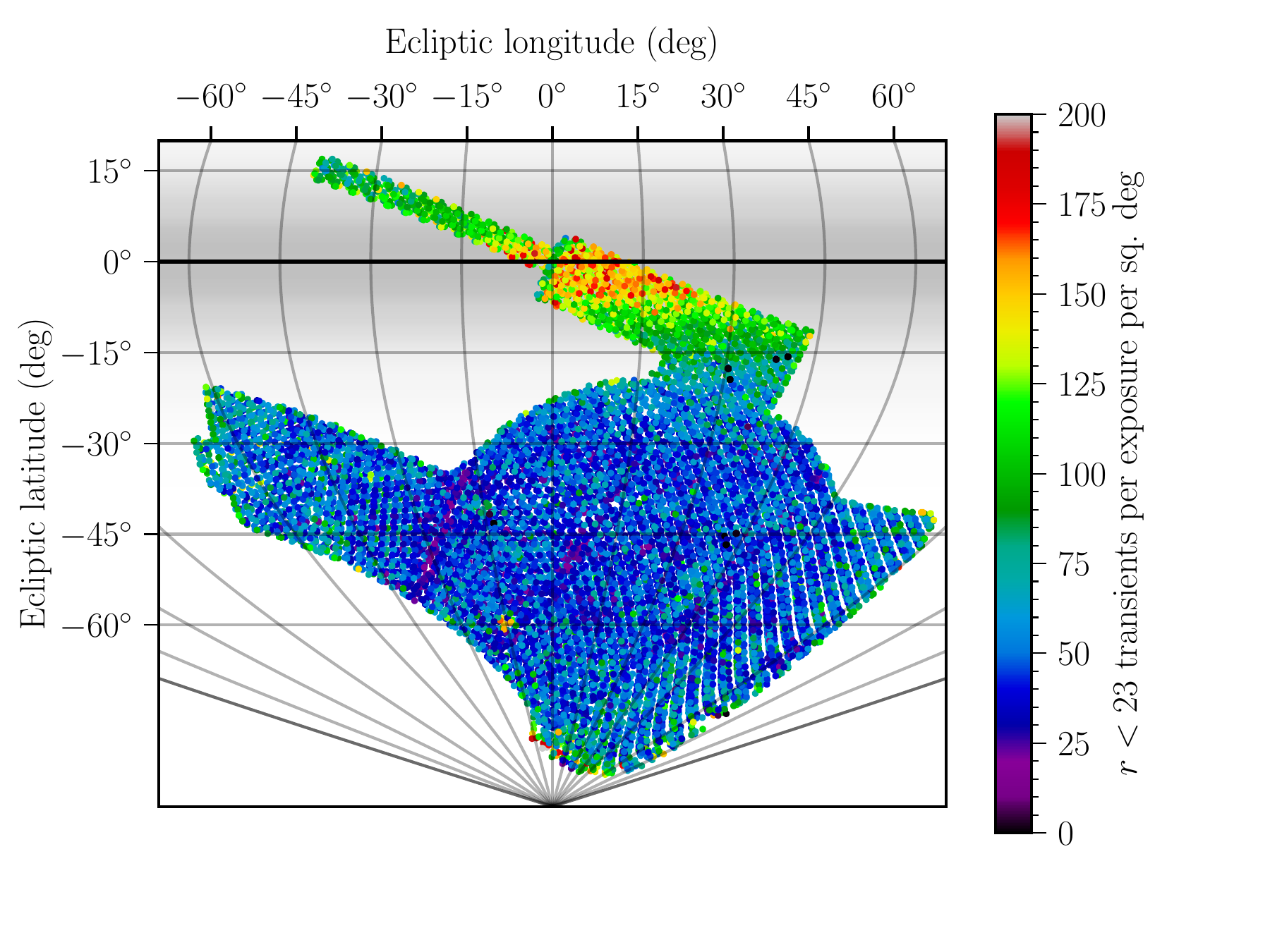}
	\caption{Sanson-Flamsteed equal-area projection of the transient map in ecliptic coordinates for the \des\ Y4 exposures, with color of each $r$-band exposure denoting the sky density of transients on this exposure. We restrict transient counts to $r<23,$ where nearly all exposures are highly complete, so that this map reflects sky density variations rather than changes in the limiting magnitude of exposures.  The density increases with ecliptic latitude, peaking at the ecliptic, being dominated by asteroids. The darker stripes running through the round central region are places where the tiling of the sphere requires us to overlap exposures more heavily.  This leads to deeper coadd images and more effective rejection of spurious transients. The shaded gray region represents the latitudinal density of the asteroid belt.\label{im:transient_map}}
\end{figure}

Another difference between the \des\ survey and those designed for TNO detection is that most of the \des\ footprint is at high ecliptic latitudes (Figure~\ref{im:transient_map}).  This will allow the \des\ to obtain stronger constraints on the high-inclination population.

Section~\ref{sec:transients} summarizes the observations and image processing procedures of the \des, details the extraction of a catalog of transient sources from these data, and evaluates the detection thresholds in the individual exposures.  Section~\ref{sec:linking} presents the algorithms used to identify TNOs from the catalog of single-epoch transients, including quantification of the rate of false-positive linkages. Section~\ref{sec:subthreshold} defines the ``sub-threshold significance'' statistic: we examine images that cover the putative object's orbit, but did not yield a detection.  A real object will be present but with signal lurking below detection threshold, while a spurious linkage will have zero signal on these unlinked images.
Section~\ref{sec:fakes} evaluates the effectiveness of the search through insertion of synthetic TNOs into the transient catalog.  The catalog of secure Y4 \des\ TNO detections is presented in Section~\ref{sec:catalog}. In Section~\ref{sec:conclusions}, we summarize and highlight the improvements to be expected from next \des\ TNO catalog release. This will make use of the full \des\ observation set, yielding improved depth and orbital parameters, but will also gain depth from several methodological upgrades.  This paper will describe only the production of the Y4 \des\ TNO catalog; analysis of its contents will be described in further publications.  While we describe here the methods that are used to determine the survey's sensitivity vs.\ TNO magnitude and orbital parameters, we will defer the full-scale application of these methods until the analysis of the final \des\ TNO catalog.  Thus we do not anticipate production or publication of the detailed completeness functions nor a survey simulator for the catalog presented herein.

For readers who are more interested in the catalog than in the details of the methods used to discover its contents, we would suggest skimming Section~\ref{sec:transients} and then skipping to Section~\ref{sec:fakes} to see the nature of the final selection function, and later sections with the contents of the catalog.  

The \des\ collaboration has experimented with methods for identifying and linking transients which differ from those presented in this paper. Many TNOs have been discovered or measured \citep{Gerdes2016,Gerdes2017,Becker2018,Khain2018,Lin2019,Khain2019} using difference imaging for transient identification \citep{Kessler2015}. The search described in this paper is the first to search the full Y4 survey in a uniform fashion. For reference, we include in this paper a table of objects that were discovered in \des\ data by the other methods, but missed in this ``Y4'' search. As will be discussed in Section~\ref{sec:catalog}, all such cases are found to be TNOs that lie outside the selection criteria of the Y4 search, i.e. the Y4 search is seen to be complete within its stated criteria.

\section{The transient catalog}
\label{sec:transients}

\subsection{Observations}
The \des\ observational strategy is fully described in \citet{Diehl2016,Diehl2018} but we summarize the relevant details here.  The ``wide'' portion of the survey aims to completely tile the 5000~deg$^2$ footprint with 
$10\times90{\rm s}$ exposures in each of the $g, r, i,$ and $z$ bands, and $6\times45 + 2\times90$~s in $Y$ band, such that the completed wide survey will comprise $\approx80,000$ DECam exposures.  \des\ observations are taken in seasons beginning in August and ending in February each year.  The wide survey exposures are interleaved with a ``deep'' survey which images 10 DECam pointings ($\approx30$~deg$^2$) at $\approx$ weekly intervals in the $griz$ bands, primarily for detection and measurement of high-redshift Type~Ia supernovae \citep{dessn}.  The Y4 TNO search reported herein was conducted only on the wide-survey images, but we have also searched much of the deep data and reported TNO detections to the Minor Planet Center (MPC) \citep{Gerdes2016,Gerdes2017,Becker2018,Lin2019,Khain2018, Khain2019}. 

Observations are scheduled with the \textsc{obstac} algorithm \citep{obstac}, which works to optimize the quality and homogeneity of the final wide-survey products in the face of variable clouds, seeing, and moonlight while balancing the needs of the deep survey for temporally regular sampling.  Importantly, part of this optimization is to avoid imaging the same part of the wide survey more than once in any filter on any single night [the goal is to spread the weather fluctuations across the footprint to homogenize the final survey products].  Occasionally \textsc{obstac} will elect to take successive exposures at the same pointing in two different filters, typically $g,r,$ and/or $i$ in dark time and $zY$ in bright conditions.  These pairs, with only 2~minute intervals between them, are not useful for detecting TNO motion.  Repeat imaging on any time interval longer than 2 minutes and shorter than one day is very rare.  When we calculate the robustness of our TNO detections against false positive linkages, we will always consider multiple detections on the same night to be fully correlated (not independent) since most sources of false positives (asteroids, variable stars, artifacts from bright sources) will tend to repeat in successive exposures.  Each \des\ exposure is processed and evaluated the next day.  Exposures failing to meet minimum standards for seeing and signal-to-noise ratio ($S/N$) are discarded, and their pointings placed back on the queue.  We do not search rejected images for TNOs.

Nominally the survey goals were to cover half of the footprint with 4 ``tilings'' per filter in the first year; cover the other half with 4 tilings in the second year; and then 2 full tilings per year for 3 succeeding years.  Variability in weather and availability of fields meant that this plan is only a rough approximation to reality.  In particular, very poor weather in year 3 put the survey behind schedule, and an additional half-season's worth of observations were allocated for a sixth year to complete the survey.  In the Y4 data under analysis here, a typical point in the footprint has been within the DECam field of view for $\approx 7$ exposures per band.  Within the field of view, 15--20\% of the sky is lost to gaps between the DECam CCDs, regions of defective CCDs, and area lost in the glare of bright sources.  Thus a typical TNO will be cleanly imaged $\approx6\times$ per filter in the Y4 data.

One other observational detail of note is that DECam has a two-blade shutter which takes $\approx 1$~s to sweep across the focal plane \citep[][section 6.1]{Flaugher2015}.  Successive exposures sweep the blades in opposite directions.
We assign all TNO detections to a time corresponding to the midpoint of the shutter-open interval in the center of the focal plane.  The true mean time of the open-shutter interval for any particular TNO's exposure may differ by up to $\approx0.5$~s in an unknown direction.  Since the TNO apparent motion in this time interval is $<1$~mas, the shutter uncertainty contributes negligibly to errors in orbit determination.

\vspace{-6pt}
\subsection{Processing and cataloging}
For the analysis presented here, we make use of two distinct types of object catalogs produced during standard \des\ processing \citep{Morganson2018}: the single-epoch (SE) produced for each individual exposure, and the multi-epoch (a.k.a. ``coadd'') catalogs produced from an image averaging all exposures.

The SE catalogs used for this search come from the ``Final Cut'' reduction of the internal ``Y4A1'' data release. This processing includes the detrending and calibration of the raw CCD data.  Artifacts such as saturated pixels close to bright stars, cosmic rays and streaks are masked, sky background template images are subtracted from the science image, and the PSF is modeled via the bright stars. Sources are detected in the images using \textsc{SExtractor} \citep{Bertin1996}, with its default $3 \times 3$-pixel filter first being applied to the image. The detection criteria are that a source needs 6 pixels ($\mathtt{DETECT\_MINAREA} = 6$) above the detection threshold $\mathtt{DETECT\_THRESH} = \mathtt{ANALYSIS\_THRESH} = 1.5$.  A variety of position, flux, and shape measurements are made on each object and recorded in the output SE catalog. 

The coadd catalogs in use here come from an internal data release labelled ``Y3A2,'' \textit{i.e.} they combine images only from the first 3 \des\ seasons.  Sources are detected on a sum of all the $r$, $i$ and $z$ images using \textsc{SExtractor} settings similar to the SE values above. The coadd catalogs and images used in this work are the same as those in the public DR1 release\footnote{\url{https://des.ncsa.illinois.edu/releases/dr1}} \citep{des_dr1}.

\begin{deluxetable}{lcc}
  \tablecaption{Overview of the methodology, showing \emph{in italics} the steps that transform the initial single-epoch detection catalogs through intermediate steps to a verified TNO catalog.  The last two columns show the number of objects at each stage and, where appropriate, the number of artificial objects that we have injected to test the pipeline efficiency.
    \label{tb:stages}}
	\tablehead{\colhead{Catalog/\textit{Processing step}} & \colhead{No. of real elements} &
        \colhead{No. of injected elements}}
	\startdata
	Single epoch detections & $7\times10^9$ & \nodata \\
	$\rightarrow$\emph{Transient identification (\textsection \ref{sec:transients})}$\rightarrow$ & & \\
	$\rightarrow$\emph{Blinded fake injection (\textsection \ref{sec:fakes_cat})} $\rightarrow$ & & \\
	Transients & $2\times10^7$  &  $3.7 \times 10^4$  \\
	$\rightarrow$\emph{Pair finding (\textsection \ref{sec:pairs})}$\rightarrow$ & & \\
	Pairs & $10^{12}$ & \nodata \\
	$\rightarrow$\emph{Triplet finding (\textsection \ref{sec:triplets})}$\rightarrow$ & & \\
	Triplets & $ 6 \times 10^{10}$ & \nodata \\
	$\rightarrow$\emph{Orbit growing (\textsection \ref{sec:grow})}$\rightarrow$ & & \\
	$\rightarrow$\emph{Fake unblinding (\textsection \ref{sec:fakes_cat})} $\rightarrow$ & & \\
	Sixlets & 1684 & 2252 \\
	$\rightarrow$\emph{Reliability cuts (\textsection \ref{sec:reliab})}$\rightarrow$ & & \\ 
	Candidates & 424 & 1727 \\ 
	$\rightarrow$\emph{Sub-threshold significance test (\textsection \ref{sec:subthreshold})}$\rightarrow$ & & \\
	Confirmed objects & 316 & \nodata
	\enddata
\end{deluxetable}

\subsection{Calibration}
The \des\ catalogs are exquisitely well calibrated to a common photometric system across the focal plane \citep{Bernstein2017phot} and across all the exposures of the survey \citep{Burke2017}.  Comparison of \des\ magnitudes to Gaia DR2 magnitudes show uniformity across the footprint to $\approx6$~mmag RMS \citep[][section 4.2]{des_dr1}.  Trailing of TNO images in the 90~s \des\ wide exposures has negligible impact on photometry: at the maximual rate of apparent motion for objects $\ge30~AU$
of $5\arcsec/\mathrm{hour}$, the object would move $0.12\arcsec$.  The second moment of the trail is then $<1\%$ of the second moment of the PSF, and the point-source calibrations should be accurate to $<0.01$~mag even for this maximal apparent motion.

Every \des\ exposure has an astrometric map from pixel coordinates to the Gaia DR1 \citep{Lindegren2016,Brown2016} celestial reference frame.  Here again we benefit from extensive \des\ calibration efforts.
\cite{Bernstein2017astro} describes the DECam astrometric model, demonstrating that all distortions induced by the telescope, instrument, and detectors are known to 3--6~milliarcsec RMS.  We apply this model to all exposures in the Y4 \des\ TNO search. The astrometric uncertainties for bright TNO detections are dominated by stochastic
displacements caused by atmospheric turbulence, at 10--15~mas RMS on typical exposures. Below we describe the estimation of this turbulence uncertainty for each exposure.  Fainter detections' astrometric uncertainties are dominated by shot noise in the centroiding of the image.

The astrometric solution includes terms for differential chromatic refraction (DCR) in the atmosphere and lateral color in the optics, which are calibrated in terms of the $g-i$ color of the source.  Similarly the photometric solution includes color terms.  The maximal amplitude of the DCR (for airmass $X<2$) and lateral color are $\approx 80$ and 40~mas per mag of $g-i$ color, respectively, for $g$-band observations, and 5 or more times smaller in other bands \citep{Bernstein2017astro}.
The $g-i$ colors of all stars are measured directly by \des, which fixes the reference frame, but the TNO apparent position will depend on its (unknown) color.
Our TNO search is executed using positions that assume a nominal color, $g-i=0.61$ (a typical stellar color).  Only after a TNO is linked can we estimate its color.  The final positions, magnitudes, and orbital parameters reported herein are calculated after refinement using the proper chromatic corrections. 

\begin{figure}[htb!]
	\centering
	\includegraphics[width=0.8\textwidth]{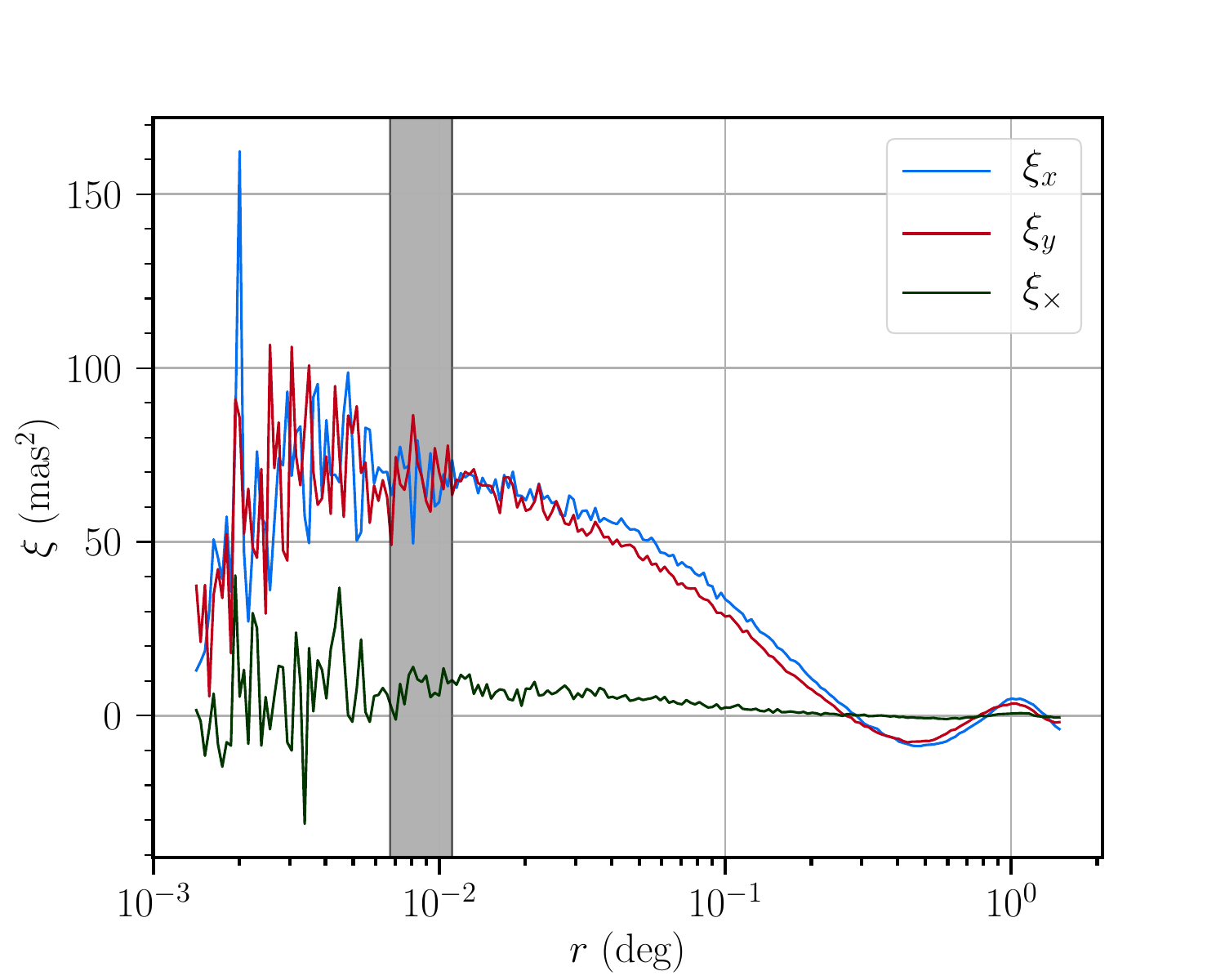}
	\caption{The correlation function $\xi(r)$ of astrometric errors for high-S/N stars separated by angle $r$.  This plot gives the average of $\xi(r)$ for all $g$ band exposures in a small region of the survey, taken across 4 years of observations. The shaded region denotes where the $\xi$ values are averaged to establish the covariance matrix of atmospheric turbulence.\label{im:turbulence}}
\end{figure}

The procedure for estimating the atmospheric turbulence contribution to astrometric errors is as follows:
\begin{enumerate}[label=\roman*.]
       \item For each high-$S/N$ star $i$ in the survey footprint, we calculate a mean position by averaging the positions predicted by the astrometric model in all \des\ exposures, as well as any position available in Gaia DR1, to produce a ``truth'' value $\mathbf{\bar x}_i$. 
	\item Restricting ourselves to the positions $\mathbf{x}_i$ measured on an individual \des\ exposure, we
find the measurement error $\Delta \mathbf{x}_i \equiv \mathbf{x}_i-\mathbf{\bar x}_i = (\Delta x_i, \Delta y_i)$ from the mean position. The displacement is measured in a local gnomonic projection, with $x$ pointing to local equatorial east and $y$ north. 
	\item A cubic polynomial function of field coordinates is fit to the $\Delta\mathbf{x}_i$ and adopted as the large-scale distortions from atmospheric and turbulent refraction for this exposure.  This fit is subtracted from the $\Delta\mathbf{x}_i.$
	\item We calculate the two-point correlation functions of astrometric error, averaging over all pairs of stars in the exposure vs.\ their separation $r$: 
	\begin{subequations}
		\begin{eqnarray}
			\xi_x(r) =& \langle \Delta x_i \Delta x_j \rangle_{|\mathbf{x}_i - \mathbf{x}_j| = r},\\
			\xi_y(r) = &\langle \Delta y_i \Delta y_j \rangle_{|\mathbf{x}_i - \mathbf{x}_j| = r},\\
			\xi_{\times}(r) = &\langle \Delta x_i \Delta y_j \rangle_{|\mathbf{x}_i - \mathbf{x}_j| = r}.
		\end{eqnarray}
	\end{subequations}
        An example of the behavior of the $\xi$'s is shown in Figure~\ref{im:turbulence}.
      \item We assign a 2-d Gaussian positional uncertainty to each detection in the exposure, with a covariance matrix given by
\begin{equation}
  \boldsymbol\Sigma_\mathrm{atm} = \begin{pmatrix} \langle \xi_x \rangle & \langle \xi_{\times} \rangle \\
    \langle \xi_{\times} \rangle & \langle \xi_y \rangle \end{pmatrix}
  \label{astsigma}
\end{equation}
where the $\xi$ values are averaged over the separation range $24\arcsec < r < 40\arcsec$ where they typically plateau.  Note that this use of the correlation function at $r>0$ removes any contribution due to shot noise of individual stellar measurements.
\end{enumerate}

\begin{figure}[htb!]
	\centering
	\includegraphics[width=0.8\textwidth]{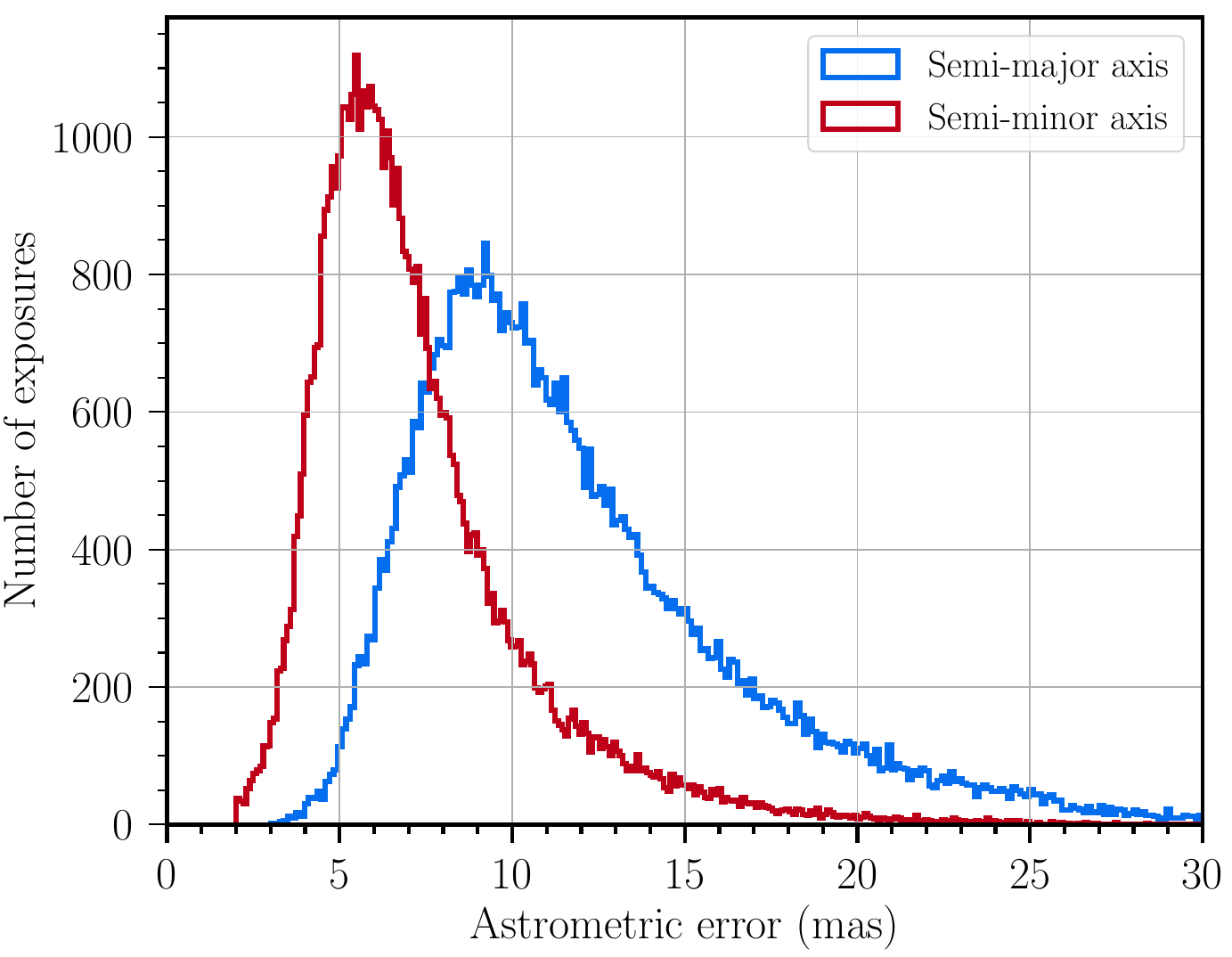}
	\caption{Distribution of the semi-major and semi-minor axes for the atmospheric turbulence error ellipses for all exposures used in the \des\ internally-released ``Y4A1'' catalog.\label{im:atm_errors}}
\end{figure}

Figure \ref{im:atm_errors} shows the histogram of atmospheric turbulence strength in the exposures.
The typical atmospheric turbulence is seen to be $\approx 10 \, \mathrm{mas}$. 
The astrometric errors are typically substantially anisotropic on most individual exposures, with $\Sigma_{xx}\ne \Sigma_{yy}$ and $\Sigma_{xy}\ne 0,$ because astrometric errors are substantially different parallel vs.\ perpendicular to the prevailing wind direction.

The final astrometric uncertainty assigned to each transient is the quadrature sum of the atmospheric turbulence ellipse in Equation~\ref{astsigma} with the circularly symmetric shot noise error in the centroid reported by \textsc{SExtractor.}

\subsection{Identifying single-night transients}
The computational burden of linking TNO's from the transient catalog scales roughly as $n^3,$ where $n$ is the density of transient detections per square degree per exposure.  It is therefore of great importance to minimize $n$ while retaining all true TNO detections in the transient catalog.  Because asteroids are indistinguishable from TNOs at the single-exposure level, the transient catalog must include all detected asteroids, which will greatly outnumber the TNOs.\footnote{Previously known asteroids could be deleted from the transient catalog, but at these magnitudes only a small fraction of asteroids are currently cataloged.} Asteroids thus form an irreducible source of false-positive TNO transient detections, and our goal, therefore, is to reduce any other spurious transient detections to a level well below the asteroid density.

We wish to identify all sources in the thousands of \des\, images which are present in a given sky location for only a single night.
We begin by matching the SE in all bands and coadd catalogs (in which each detection has information from all bands) by sky coordinates.
All detections in a small region are projected into a tangent plane and then grouped with a kD-tree \citep{Maneewongvatana1999} friends-of-friends (FoF) algorithm that
links detections within $0\farcs5$ of each other.

For each group of SE detections, we calculate the following quantities:
\begin{itemize}
	\item Whether or not it is matched to a coadd detection, $\mathtt{COADD} = 1,0$;
	\item How far apart in time are the first and the last SE detections of the match group, $\Delta t \equiv t_\mathrm{last} - t_\mathrm{first}$;
	\item How much fainter or brighter the SE detection is compared to the coadd detection in the same band (when present), $\Delta m \equiv m_\mathrm{SE} - m_\mathrm{coadd}$.
\end{itemize}
If the SE detection is a detection of a solar-system object, we do not expect any flux to be present at this location on the other exposures contributing to the coadd taken more than a few hours away from the SE detection.  If there are $K$ exposures in the coadd, then the averaging process will reduce the apparent flux of the source in the coadd by a factor $1/K$, leaving the coadd source with a magnitude fainter than the SE value by $2.5 \log_{10} K \ge 0.75$~mag for $K\ge 2$.  The coadd source should thus either be absent or significantly fainter than the SE source. We therefore implement the following cut to retain potential solar system measurements while eliminating many variable fixed (stellar) sources:
\begin{enumerate}
	\item $\mathtt{COADD} = 1$, $\Delta t < 2 \, \mathrm{days}$, $\Delta m \leq -0.4$; \textrm{OR}
	\item $\mathtt{COADD} = 0$, $\Delta t < 2 \, \mathrm{days}$.
\end{enumerate}  

\begin{figure}[htb!]
	\centering
	\includegraphics[width=0.8\textwidth]{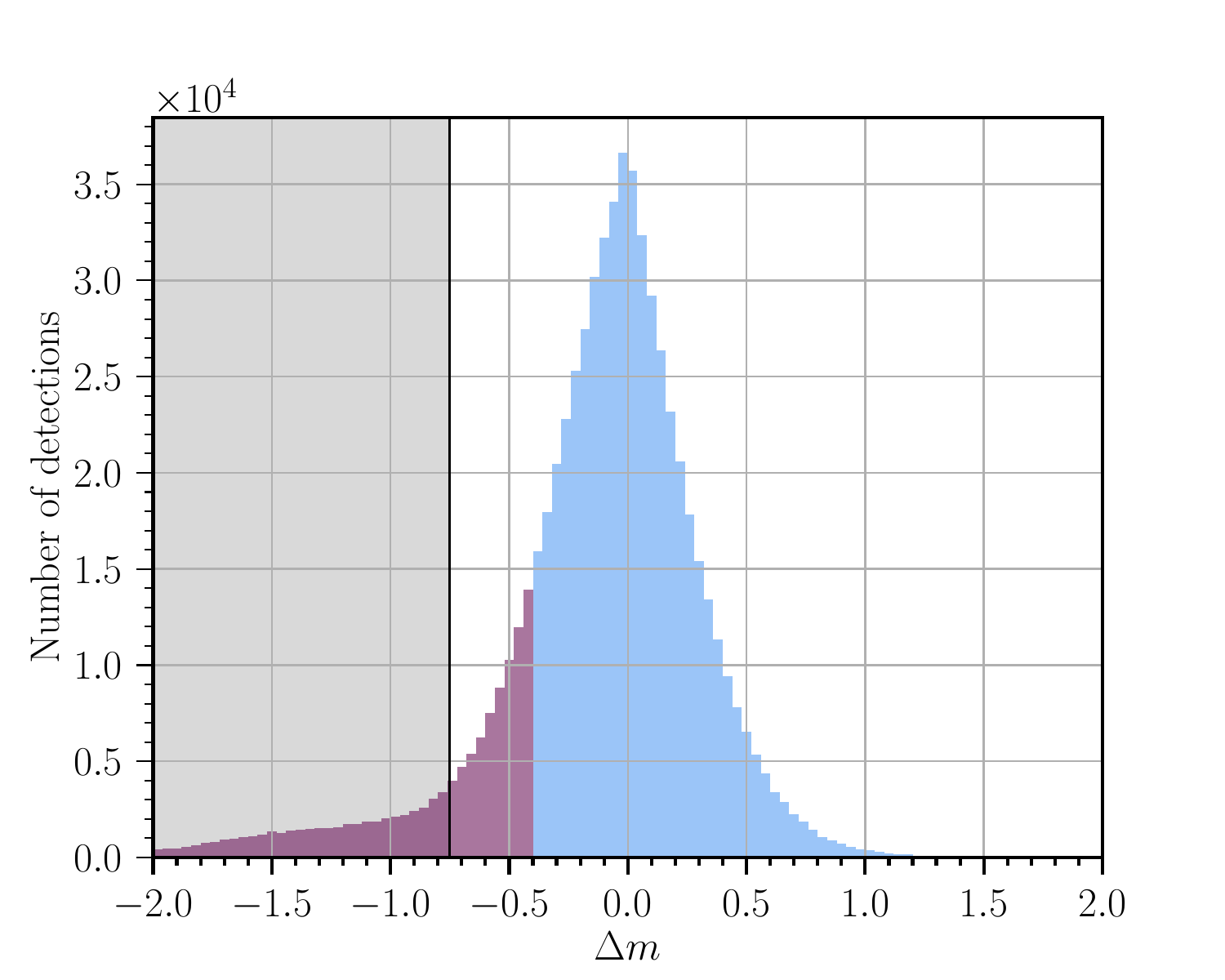}
	\caption{The distribution of $\Delta m$, the difference in magnitude between single-night detections and the magnitude of the matched object in the coadd frame, is shown for detections in
$25 \deg^2$ of the survey. We posit this to be the sum of a population with symmetric distribution about $\Delta m=0$ caused by noise and variability of static sources, plus a tail of highly negative $\Delta m$ values containing moving objects (peaking at $\Delta m \approx -1.5$). True single-night transients are expected to have, on average, $\Delta m  < -0.75$ (shaded region), because the single-night flux will be reduced by the number of (zero-flux) other exposure's data averaged together.
All objects in the red region at $\Delta m < -0.4$ are included in the transient catalog, with this cut selected to give us high probability of capturing all of the true single-night transients.
\label{im:delta_m}}
\end{figure}

Note that $\Delta t <2$~days is by itself an insufficient condition, since it would include many faint sources that are pushed above detection thresholds on only 1--2 exposures because of intrinsic variability or noise fluctuations.  The veto by the more sensitive coadd images solves this issue, with the $\Delta m \le -0.4$ threshold estimated empirically to include nearly all truly moving sources (TNOs), as illustrated in Figure~\ref{im:delta_m}.

Unlike \cite{Goldstein2015} or \cite{Lin2018hsc}, we do not make cuts on object sizes or other morphological features of detections, as 
these measures are too noisy to be useful discriminants of non-stellar artifacts from TNOs for sources at the threshold of detectability.  We are willing to accept higher false-positive rates in order to keep the lowest flux threshold for true TNOs.

\begin{figure}[tb]
	\centering
	\includegraphics[width=0.9\textwidth]{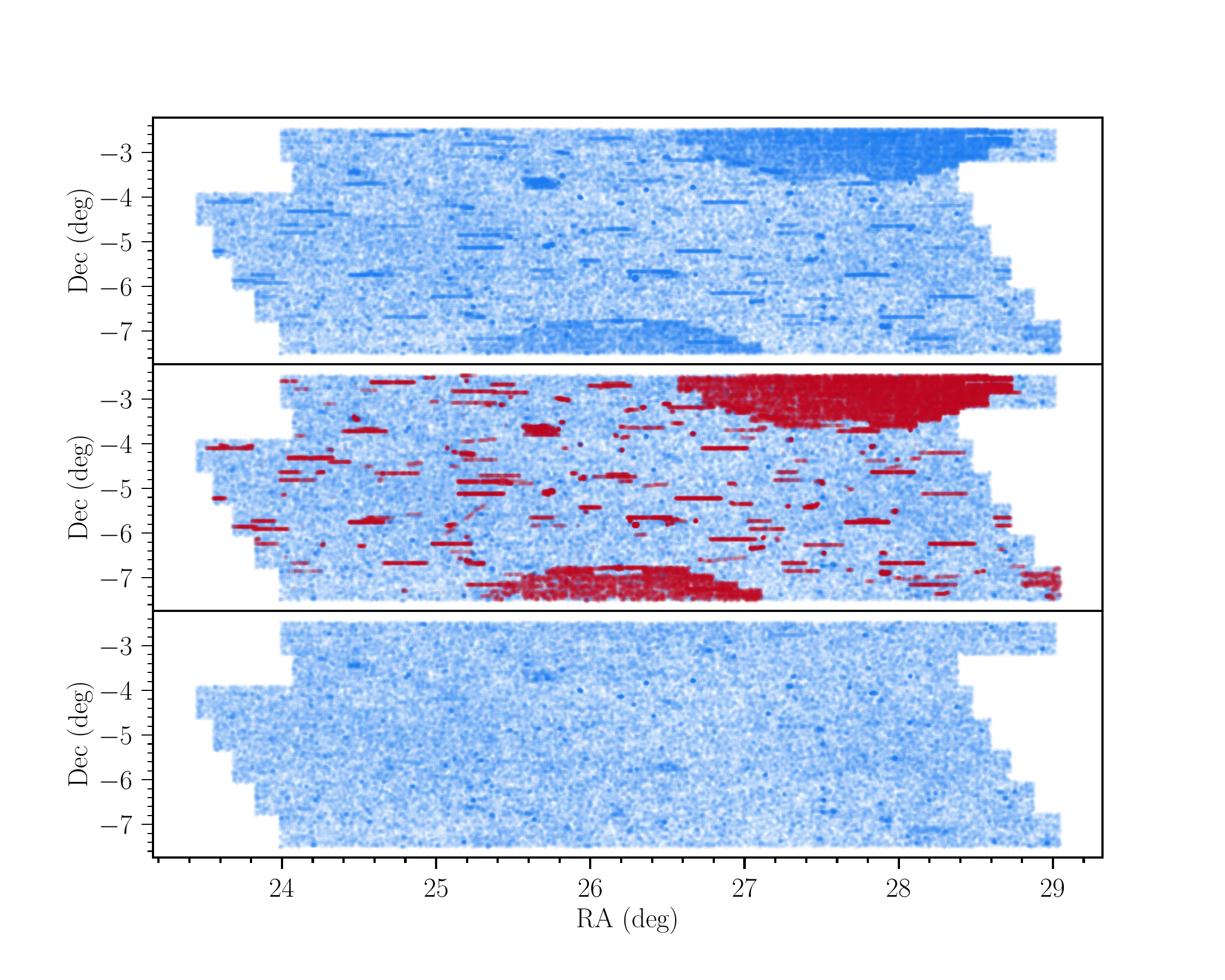}
	\caption{The three stages of the transient catalog. From top to bottom: 1) the catalog before the cleaning process; 2) identification of the coherent structures; 3) coherent structures are removed, showing the output catalog. Note that there are still some coherent structures remaining, but these are now a small minority of the total transient catalog.\label{im:cleanup}}
\end{figure}

\begin{figure}[tb]
	\centering
	\includegraphics[width=0.9\textwidth]{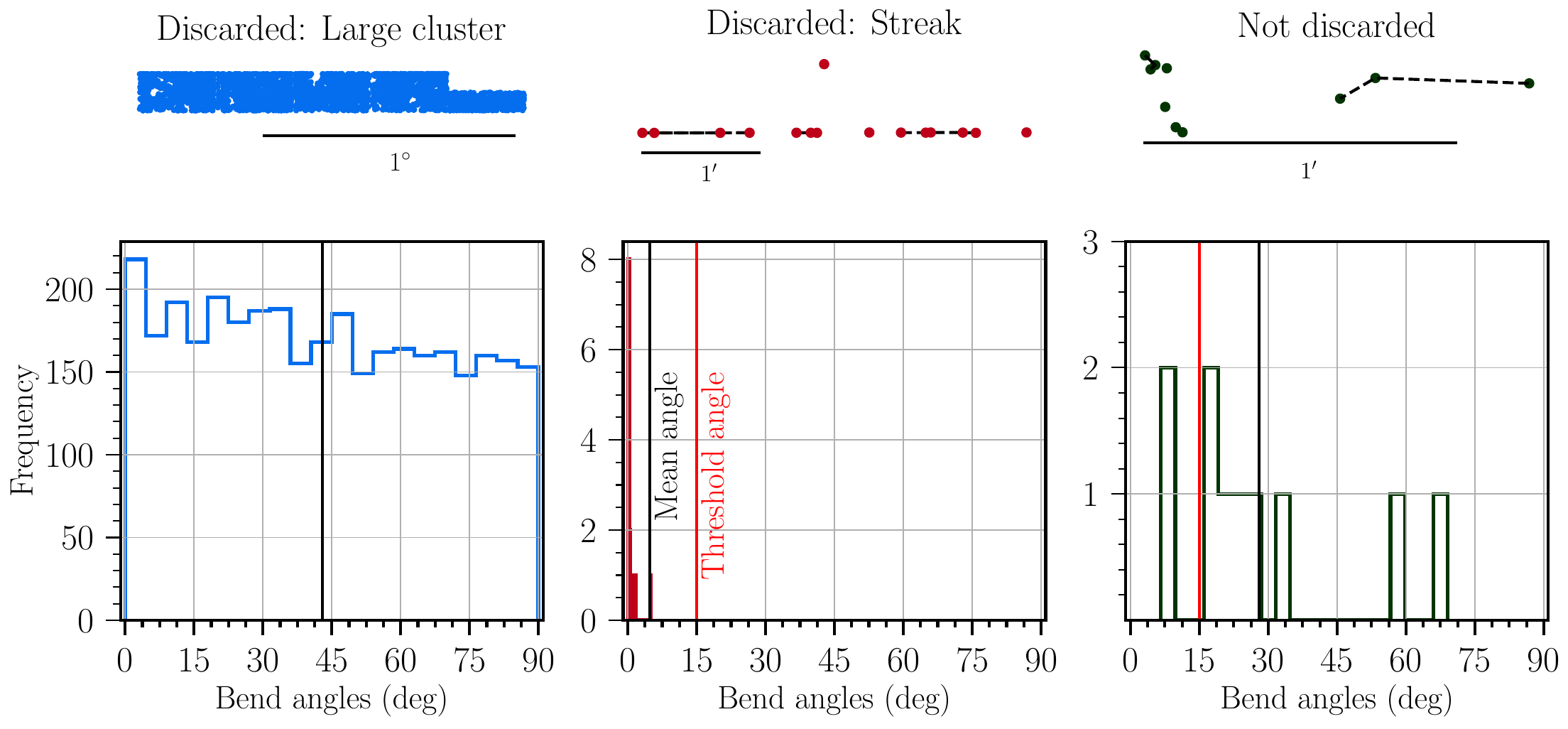}
	\caption{The bend angle and clustering evaluation. In the first column, the transients from an exposure with astrometric misalignment (top), showing randomly distributed bend angles (bottom). In the second column, a streak (with one spurious match) identified in the cleaning process (top), and the bend angle histogram (bottom) showing values close to zero. In the third column, a randomly matched group (top), with no tendency in the bend angle histogram (bottom). In the last two scatter plots, the black dashed lines show some of the connected nearest neighbors, which define (visually) the bend angle.\label{im:angles}}
\end{figure}

The first of Figures~\ref{im:cleanup} maps the sky locations of transient detections, revealing many coherent structures that cannot be produced by TNOs, asteroids, or other true celestial moving objects.  
These structures correspond to image artifacts that were not successfully masked by the SE pipeline, such as meteor/satellite/airplane streaks, bad pixel columns, and artifacts from detections around bright sources.  There are also some exposures with incorrect astrometric solutions, 
creating large numbers of spurious sources which do not match sky coordinates of other exposures' detections.
We find empirically that the following steps applied to each exposure serve to reduce the number of spurious transient detections to level comparable to or below the irreducible asteroid density:
\begin{enumerate}[label=\roman*.]
\item The detections are clustered with FoF linking length of $60\arcsec$;
\item Groups with more than 20 detections are removed from the catalog, as these correspond to entire CCDs or even exposures being identified as transient sources;
\item Groups that have between 10 and 20 detections could come from either accidental matching of true transients, or from coherent structures like unmasked streaks or arcs around bright stars; 
\item To remove the latter, we test for a tendency of the points to lie along a line or curve:
we evaluate the ``bend angle'', that is, the angle between the lines connecting each detection on this group to its two nearest neighbors. These angles are clipped to the range $0\degr$--$90\degr$.  If the average angle of all detections in the cluster is less than $15\degr$, these detections are also removed from the catalog;
\item We also remove transients within 30 pixels of any edge of a CCD, since source images cut off at CCD edges have mis-measured centroids that cause spurious mis-matches with their corresponding coadd sources.
\end{enumerate} 
Figure \ref{im:angles} illustrates the three cases described here. The second panel of Figure \ref{im:cleanup} shows the identified structures in a region of the survey, and the last panel shows the output, and final, transient catalog. Some spurious structures remain, but since the number of these is now well below the number of asteroids, there is little to be gained by further cuts on the transient catalog. 

There are $\approx7\times10^9$ single-epoch detections in the 60,000 Y4 exposures.  Of these, $2\times10^7$ are designated as transients (see table \ref{tb:stages}).

\subsection{Transient density}
Figure \ref{im:transient_map} presents a map of the density of bright ($r < 23$) transients for each \des\  Y4 $r$ band exposure. At these magnitudes, most exposures are virtually complete (that is, the magnitudes of 50\% completeness of each exposure is on average larger than this limit, see section \ref{sec:m50} for a discussion on the completeness estimates), and the drop in density from exposures with $\sim 200 \, \mathrm{transients}\,\mathrm{deg}^{-2}$ close to the ecliptic plane to $< 100 \, \mathrm{transients}\,\mathrm{deg}^{-2}$ far from it suggests that asteroids are comparable in the catalog to other astrophysical transients and artifacts. 

\begin{figure}[htb!]
	\centering
	\includegraphics[width=0.49\textwidth]{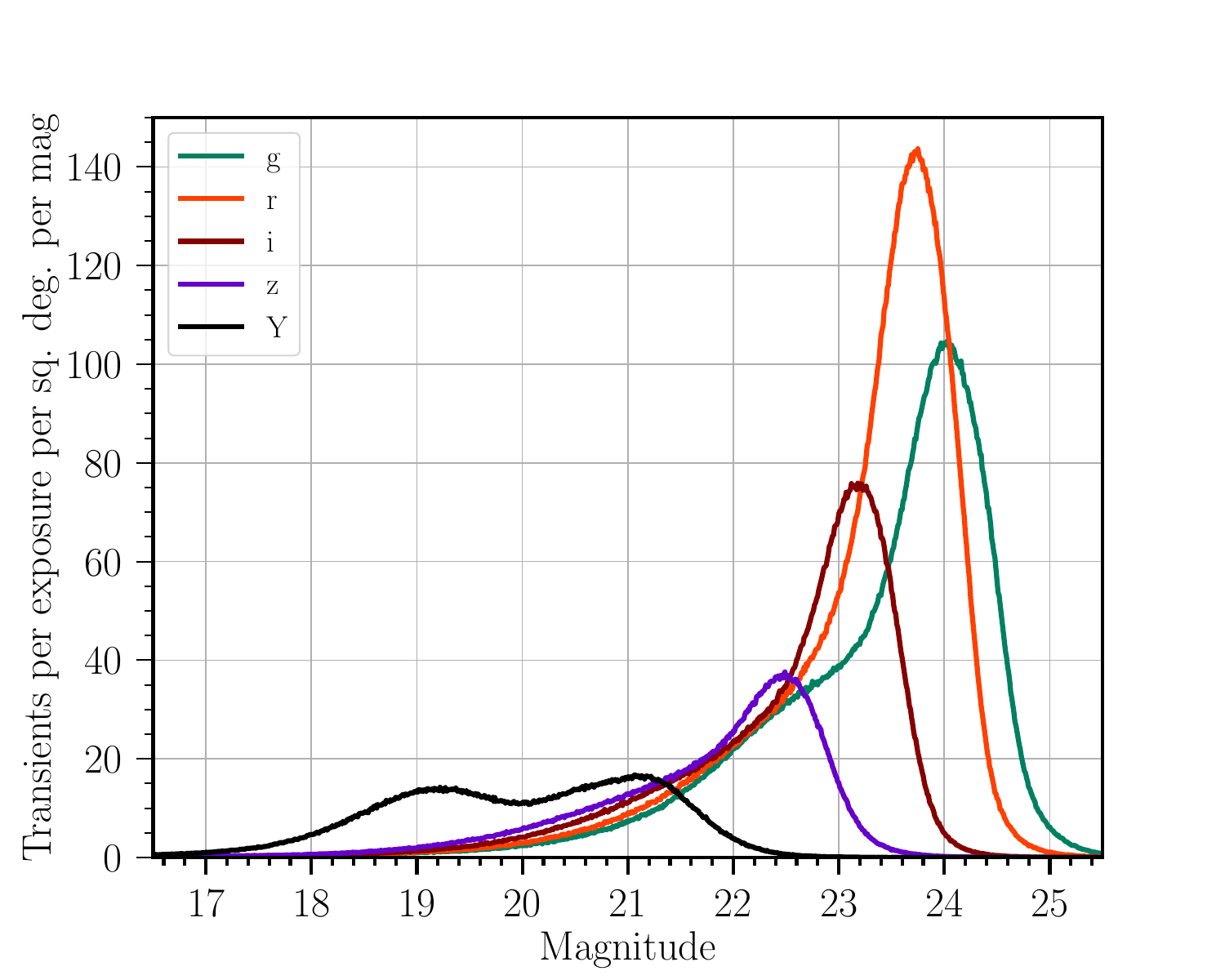}
	\includegraphics[width=0.49\textwidth]{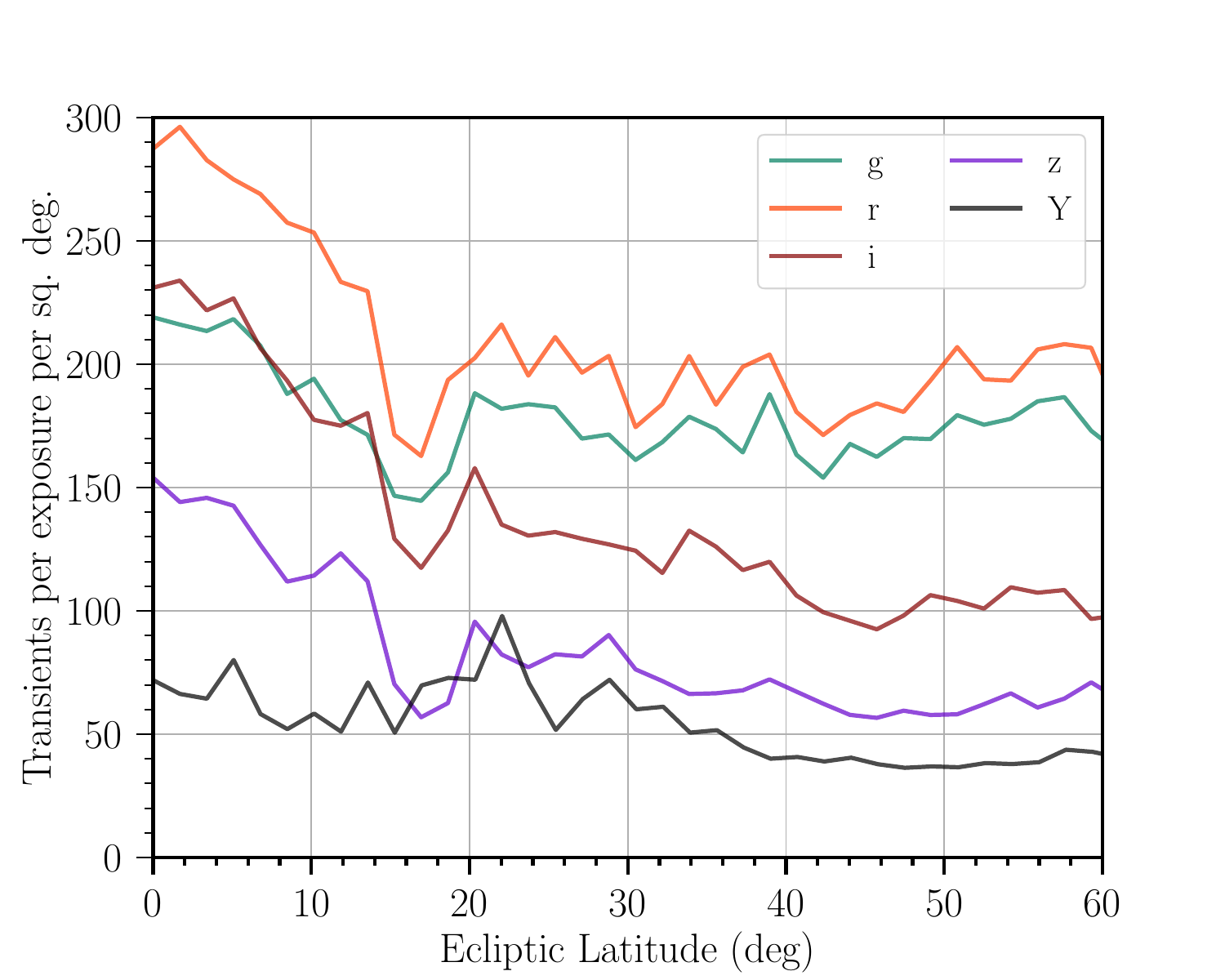}
	\includegraphics[width=0.49\textwidth]{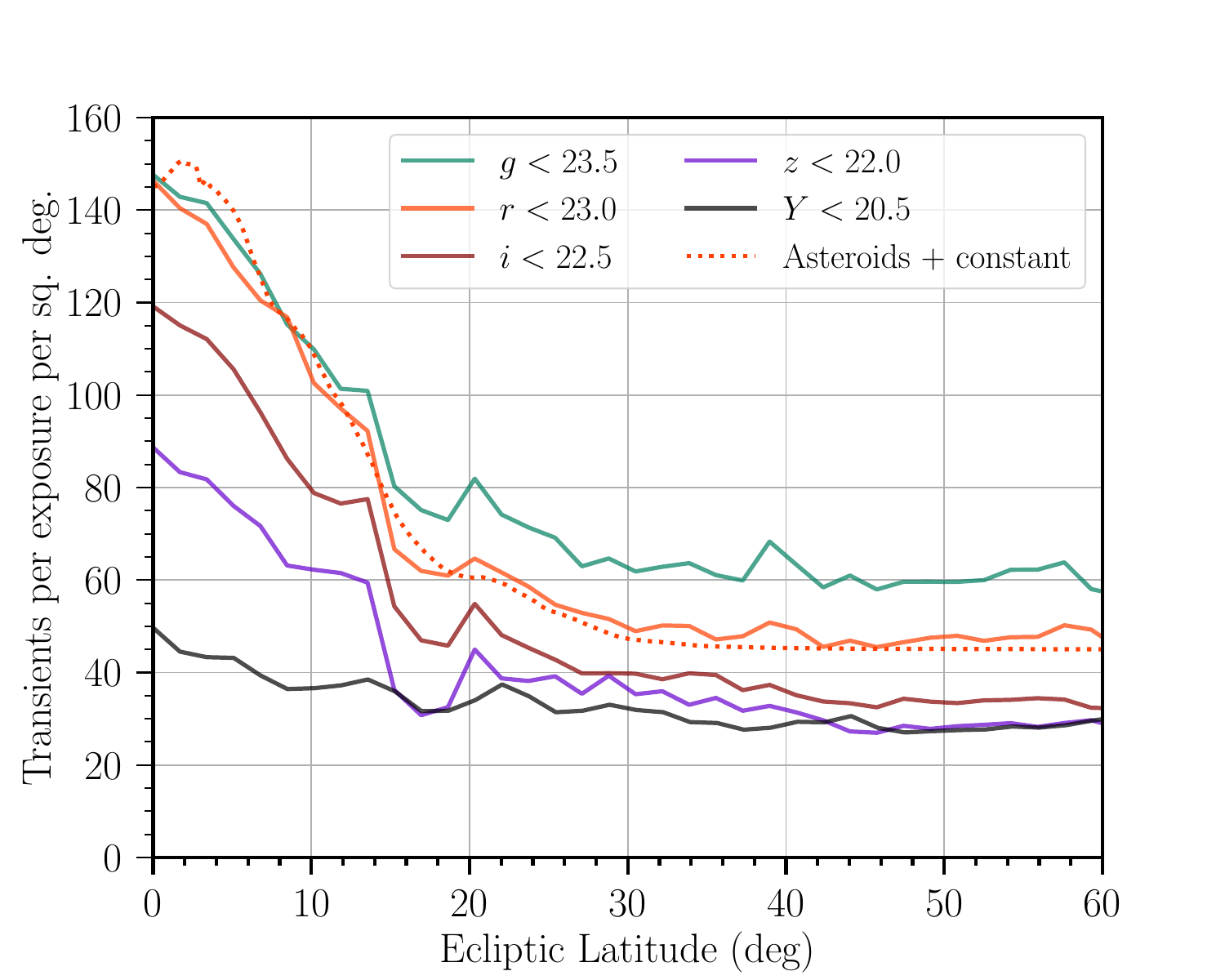}
	\includegraphics[width=0.49\textwidth]{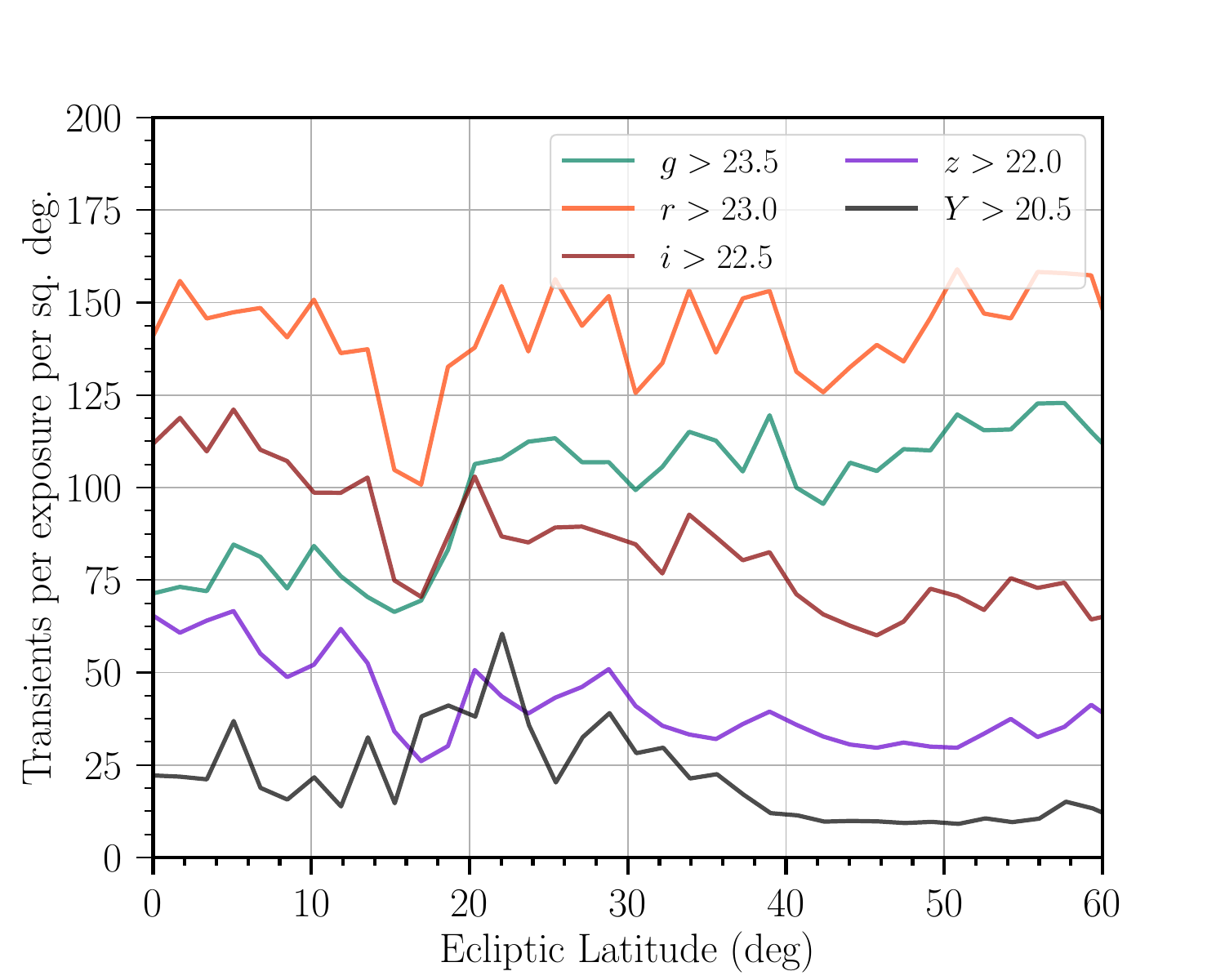}
	\caption{\emph{Upper left:} Histogram of magnitudes for all 22 million \des\ Y4 transients. \emph{Upper right:} Average transient density as a function of ecliptic latitude for all exposures. \emph{Lower left:} Bright (i.e. on the limit where most exposures are complete) transient density as a function of ecliptic latitude (similar to Figure \ref{im:transient_map}). The orange dotted curve corresponds to the latitudinal density of the asteroid belt plus a background level. \emph{Lower right:} Faint transient density as a function of ecliptic latitude.\label{im:density}}
\end{figure}

This density is consistent with the sky density of 210 asteroids
$\mathrm{deg}^{-2}$ brighter than $m_R = 23$ reported by SKADS
\citep{Gladman2009} for observations within a few degrees of
opposition.
\des\ may encounter fewer objects given that observations are not
typically at opposition. We note that this density is consistent with the latitudinal asteroid density, computed from transforming the inclination distribution for all asteroids in the Minor Planet Center\footnote{\url{https://www.minorplanetcenter.net}} to an ecliptic latitudinal distribution following \cite{Brown2001}, plus a background level of 45 transients $\deg^{-2}$ independent of ecliptic latitude, as presented in the lower left panel of Figure \ref{im:density}. 

We do not investigate the source of this background.  They are not
supernovae: simulations of \des\ supernovae transient detections
\citep{Kessler2019}, using supernovae rates from \cite{Jones2018},
that we expect only order unity supernovae (Ia and Core Collapse) per
exposure to have $r<23$ and yet appear in only one night's exposure.

The upper right-hand panel of Figure \ref{im:density} shows the average transient density in all exposures for all transients and, while the lower density far from the ecliptic is still present in the $griz$ bands, the density plateaus at $\sim 200$ per square degree per exposure ($r$ band), defining the minimum false-positive rate for our search.  The plateau at $\sim 150 \, \mathrm{transients}\,\mathrm{deg}^{-2}$ on the faint end corresponds to astrophysical transients (for example, stellar outbursts and supernovae that passed the $\Delta t$ cut, as well as high-inclination asteroids), faint sources detected only once that passed the $\Delta m$ cut, as well as image artifacts (e.g. cosmic rays, unmasked streaks). Image inspection on a randomly selected subset of false positives (identified in false TNO linkages, see section \ref{sec:subthreshold}) indicates that most of the background comes from unmasked image artifacts. While the single-night transient catalog is not pure, the level of false positives is low enough that the search is feasible. 

\subsection{Detection efficiency}
\label{sec:m50}
\begin{figure}[htb!]
	\centering
	\includegraphics[width=0.47\textwidth]{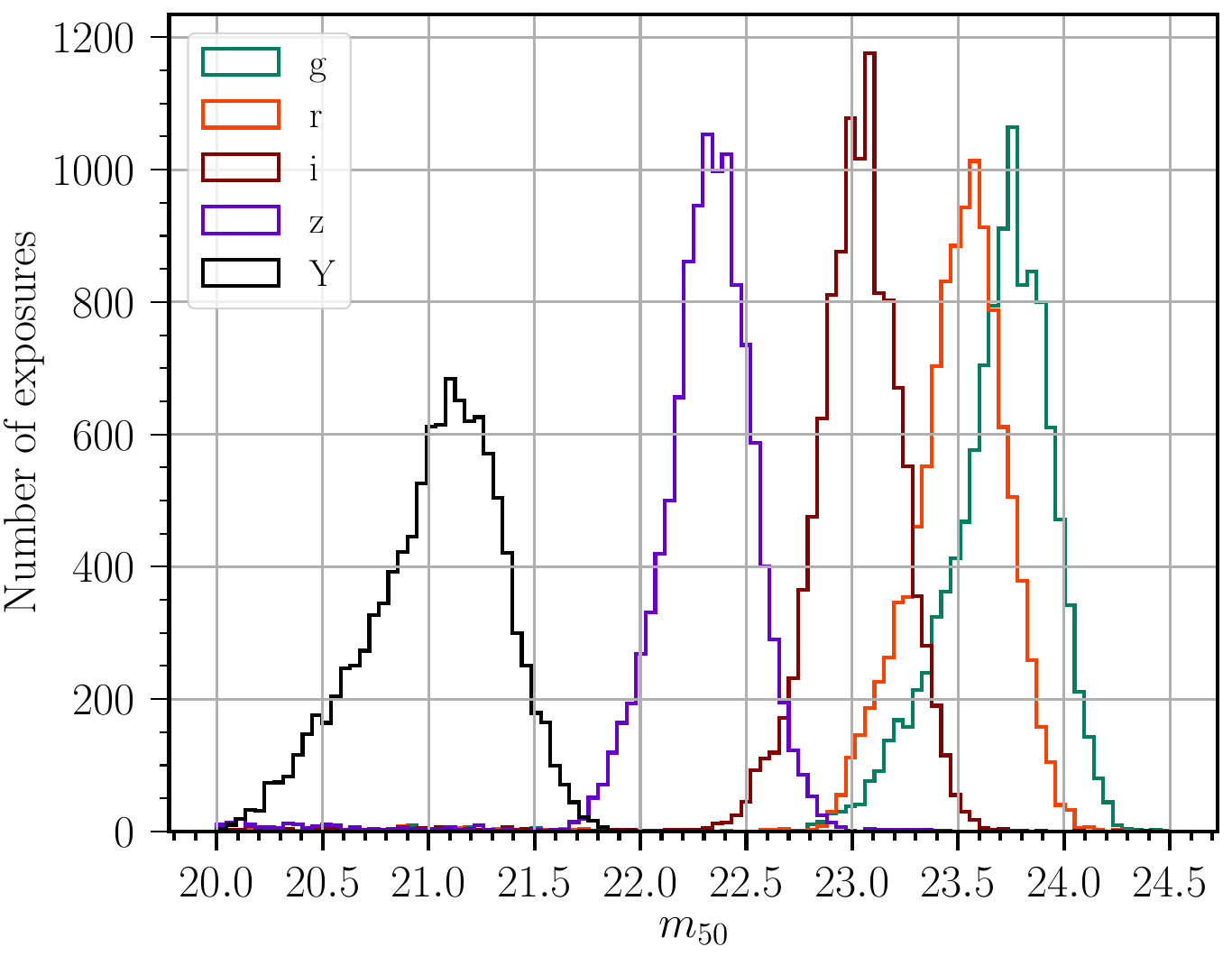}
	\includegraphics[width=0.51\textwidth]{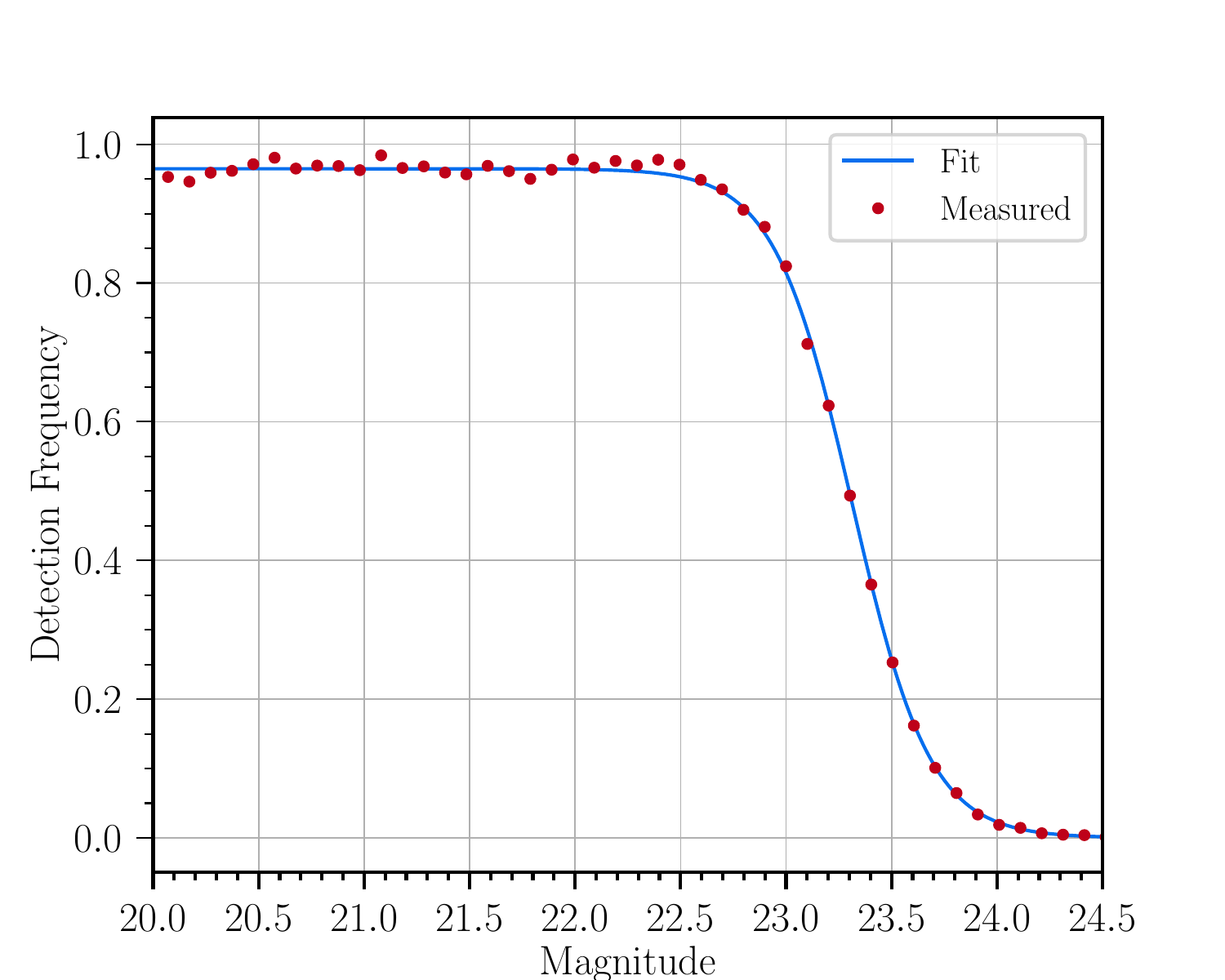}
	\caption{\emph{Left:} Histogram of the magnitude of 50\% completeness, $m_{50}$, per band of the exposures used in the \des\ Y4 TNO search, as defined in section \ref{sec:m50}. \emph{Right:} Example of the logit fit for an $i$ band exposure. The measured parameters are $(m_{50}, c, k) = (23.31, 0.96, 5.40)$. \label{im:m50}}
\end{figure}

To characterize the detection efficiency for point sources in a given exposure, we search the coadd catalogs for unresolved sources overlapping the exposure's footprint. Here, unresolved sources are defined as those with $|\mathtt{SPREAD\_MODEL\_I}| < 0.003$ \citep[see section 8.1 of][]{Drlica-Wagner2018}. We then note each coadd source's magnitude in the band of the exposure, and record whether the source was detected in the SE processing of that exposure.  This yields a list ${m_{{\rm det},i}}$ of the ``true'' (coadd) magnitudes of SE-detected point sources and another list ${m_{{\rm non},j}}$ of non-detected point sources in the exposure.  If we posit a probability of detection $p(m)$ for this exposure, then the total probability of the observed outcome is
\begin{equation}
  p_{\rm tot} = \prod_i p\left( m_{{\rm det},i}\right)  \prod_j \left[1- p\left( m_{{\rm non},j}\right) \right].
\end{equation}
We posit that the completeness function for the exposure takes the form of a logit function
\begin{equation}
  p(m) = \frac{c}{1 + \exp(k(m - m_{50}))},
  \label{eq:completeness}
\end{equation}
and we adjust the parameters $m_{50}$, $c,$ and $k$ to maximize $p_{\rm tot}.$ These then define the completeness function we adopt for this exposure.  Figure~\ref{im:m50} shows the distribution of $m_{50}$ values for the exposures included in the Y4 wide-survey TNO search, as well as an example of the fit in one exposure.  

\section{Linking TNO detections in the transient catalog}
\label{sec:linking}

We work in a modified version of the \citet[hereafter BK]{Bernstein2000} formalism, which we present briefly here.
All sky coordinates and state vectors are given in the ICRS, with origin at the solar system barycenter as tabulated by the JPL Horizons ephemerides\footnote{\url{http://ssd.jpl.nasa.gov}}.  Barycentric orbital elements are calculated from state vectors using the total mass of the Sun plus all 8 planetary systems as a central mass.  During orbit fitting, gravitational accelerations are calculated using 5 gravitating masses: one at the location of the Sun (with the mass of the terrestrial planets included), and one at each of the giant-planet barycenters, using the DE-430 ephemerides. 

Observed angular positions ($\ra$ and $\dec$) are projected to the gnomonic coordinates $(\theta_x, \theta_y)$ for the active patch
(see \citetalias{Bernstein2000} for the proper transformation equations). The angular location of an orbit with phase space vector $\mathbf{P} = \{x_0, y_0, z_0, \dot{x}_0, \dot{y}_0, \dot{z}_0 \}$ at time $t_0$  is written
\begin{subequations}
	\begin{eqnarray}
		\theta_x(t)  = & \frac{\alpha + \dot\alpha (t - t_0) + \gamma g_x(t) - \gamma x_{\E} (t)}{1 + \dot\gamma (t - t_0) + \gamma g_z(t) - \gamma z_\E(t)}, \label{eq:theta_x} \\
		\theta_y(t)  = &\frac{\beta + \dot\beta (t - t_0) + \gamma g_y(t) - \gamma y_{\E} (t)}{1 + \dot\gamma (t - t_0) + \gamma g_z(t) - \gamma z_\E(t)}. \label{eq:theta_y}
	\end{eqnarray}
\end{subequations}
Here, $\mathbf{x}_\E = (x_\E, y_\E, z_\E)$ is the location of the observatory in a barycentric reference frame relative to the origin $\mathbf{x}_0$.  
$\mathbf{g} = (g_x, g_y, g_z)$ is the gravitational perturbation, defined by
\begin{subequations}
	\begin{eqnarray}
		\mathbf{g}(t_0)  & =  \dot{\mathbf{g}}(t_0) =  0, \\
		\ddot{\mathbf{g}}(t)  & =  - \sum_i G M_i \frac{\mathbf{x}(t) - \mathbf{x}_i (t)}{|\mathbf{x}(t) - \mathbf{x}_i(t)|^3},
	\end{eqnarray}
\end{subequations}
where sum is done over the other bodies in the solar system, and
\begin{subequations}
	\begin{eqnarray}
		\alpha \equiv \frac{x_0}{z_0}, \quad  \beta \equiv \frac{y_0}{z_0}, \quad  \gamma \equiv \frac{1}{z_0}, \\ 
		\dot\alpha \equiv \frac{\dot{x}_0}{z_0}, \quad  \dot{\beta} \equiv \frac{\dot{y}_0}{z_0}, \quad  \dot{\gamma} \equiv \frac{\dot{z}_0}{z_0}.
	\end{eqnarray}
\end{subequations}
In this basis, the kinetic energy of a bound orbit is limited by
\begin{equation}
	\dot\alpha^2 + \dot\beta^2 + \dot\gamma^2 \leq 2 G M_{\sun} \gamma^3. \label{eq:bound}
\end{equation}
Here, we use the solar mass instead of the mass used to determine the barycenter, as, for our purposes, the distinction is insignificant. 

The recovery of TNOs from the full transient catalog proceeds by covering the full \des\ footprint with overlapping circular ``patches'' of radius 3.5\arcdeg\ having centers $(\ra_i,\dec_i)$.  We also divide the full search range of heliocentric distances $30<d<2500$~AU into bins of inverse distance 
$\gamma\equiv 1/d$ such that $\gamma_j-\delta\gamma_j \le \gamma \le \gamma_j+\delta\gamma_j.$ The search proceeds as
\begin{enumerate}[label=\arabic*.]
\item For each patch $i$:
  \begin{enumerate}[label*=\roman*.]
  \item Define a gnomonic projection of sky coordinates about the pole $(\ra_i,\dec_i)$, 
    to coordinates  $\boldsymbol\theta = (\theta_x, \theta_y)$ with axes directed to ecliptic east and north, respectively.
  \item For each distance bin $j$:
    \begin{enumerate}[label*=\arabic*.]
    \item For each \des\ observing season $k=1,2,3,4:$
      \begin{enumerate}[label*=\roman*.]
      \item Define a reference time $t_0$ at the midpoint of the observing season, and define an inertial Cartesian reference system with the $x$ and $y$ axes parallel to the $\theta_x,\theta_y$ directions, and coordinate origin $\mathbf{x}_0$ at the location of the observatory at $t_0.$
      \item Identify all pairs of transients within 90 days of each other in season $k$ which are compatible with a common bound orbit in distance bin $j$ and with observed position at time $t_0$ lying within patch $i$.
      \item For each pair, identify all transients within an additional 90 days of the later member that
        form a triplet compatible with a bound TNO orbit within the distance bin.
      \item ``Grow'' each triplet into candidate $n$-lets by iterating the following process: fit an orbit to the $n$ associated transients, using a prior that favors bound orbits within the distance bin (Equations \ref{eq:distprior} and \ref{eq:bindprior}).  Then find all potential additional transients (in any season) whose position is consistent with the fitted orbit.  Each such transient spawns a new $(n+1)$-let. Multiple output orbits can result from a single triplet.  Discard any $n$-let whose transients are a subset of another $n$-let, or whose transients fall on $N_{\rm unique}<5$ distinct observing nights.
      \end{enumerate}
    \end{enumerate}
  \item Merge all of the candidate TNO $n$-lets from this patch by first re-fitting the orbit with the distance priors removed, then ``growing'' to incorporate any additional transients that fit the orbit, and then removing any that duplicate or are subsets of other $n$-lets.  Each is now a candidate TNO orbit, and cuts are applied as described below to guard against false-positive linkages. 
    \end{enumerate}
  \item Merge the detections from all patches by removing duplicates.
  \end{enumerate}

These steps are described in more detail below.
Note no requirements are placed on magnitude agreement when linking transients, since (a) the majority of transients will be near the detection thresholds, and such a cut will not be very effective; (b) we do not want to exclude TNOs with high-amplitude light curves, and (c) our search is conducted in multiple filters, and we do not want to bias our detections to particular colors of TNOs by demanding a particular difference between magnitudes in distinct bands.

\subsection{Finding pairs}
\label{sec:pairs}

For short arcs of distant objects, we can neglect the gravitational perturbation, and the effect of $\dot\gamma$ on observed position is highly suppressed. Rewriting equations \ref{eq:theta_x} and \ref{eq:theta_y} with $g,\dot\gamma = 0$,
\begin{subequations}
	\begin{eqnarray}
		\alpha + \dot\alpha(t - t_0) = & (1 - \gamma z_\E)\theta_x + \gamma x_\E, \label{eq:parallax_x}\\
		\beta + \dot\beta(t - t_0) = & (1 - \gamma z_\E)\theta_y + \gamma y_\E \label{eq:parallax_y}.
	\end{eqnarray}
\end{subequations}
If we assume a
distance $d = 1/\gamma,$ this allows transformation of the $(\theta_x,\theta_y)$ coordinates into an $(\alpha,\beta)$ system where the motion is linear. \cite{Holman2018} exploit this result for linking of tracklets, but we must link individual detections. From equation \ref{eq:bound}, it becomes clear that after a time $\Delta t$ an object in a bound orbit will be in a circle in the $(\alpha,\beta)$ plane of radius 
\begin{equation} 
	r_2(\gamma, \Delta t) = \sqrt{2GM_{\sun} \gamma^3}\Delta t\label{eq:search_radius} 
\end{equation}
 centered around the first detection. 

We start by selecting all exposures from season $k$ which could potentially contain a TNO with $\gamma$ within the selected bin and $(\alpha, \beta)$ within the $3.5\degr$ patch radius. 
We apply the transformations from equations \ref{eq:parallax_x} and \ref{eq:parallax_y} for $\gamma_j$
to all transients detected in each exposure, and each set of resulting positions is used to constructed a kD tree. Trees for exposure pairs that are up to 90 days apart from each other are then searched for pairs of detections with a search radius as defined in equation \ref{eq:search_radius}. This process is repeated at the edges of the distance bin ($\gamma_j \pm \delta\gamma_j$), resulting in the final list of pairs for the bin $j$.

The expected number of pairs of unassociated transients grows with the area of the search circles and the time interval between pairs.
The mean number of such false-positive pairs between an exposure $\mu$ and all later exposure $\nu$ is 
\begin{equation}
	N_{2,\mu}(\gamma) = \sum_{\nu > \mu} 2 \pi G M_{\sun} \gamma^3 \Delta t_{\mu\nu}^2 n_\mu n_\nu A_\mathrm{DECam}
\label{paircount}
\end{equation}
where $n_{\mu}$ is the sky density of transients in exposure $\mu,$ $\Delta t_{\mu\nu}$ is the time interval between these two, and $A_\mathrm{DECam}$ is the imaging area of DECam. (This equation ignores the case where the search circle only partially overlaps exposure $\nu.$)
The total number of pairs expected in the search is $N_2 = \sum_\mu N_{2,\mu}.$  The Y4 search yields $\approx10^{12}$ pairs (Table \ref{tb:stages}).

\subsection{Finding triplets}
\label{sec:triplets}
With a pair of detections we can compute $\dot\alpha$ and $\dot\beta$ as a function of $\gamma$ and $\dot\gamma$.
A pair's \emph{nominal} expected position at future exposure at time $t$ is determined by equations \ref{eq:theta_x} and \ref{eq:theta_y} with $\alpha, \beta$, $\dot\alpha$ and $\dot\beta$ computed for the bin center $\gamma_j$ and $\dot\gamma = 0$.  We further assume that the future position is approximately linear in $(\gamma-\gamma_j)$ (``parallax'' axis) and $\dot\gamma$ (``binding'' axis) as long as we remain within the
distance bin range $\gamma_j\pm \delta\gamma_j$ and at $\dot\gamma$ small enough to maintain a bound orbit.

To compute the line of variation in $(\theta_x,\theta_y)$ due to $\gamma$ deviations (parallax), we calculate the positions for orbits corresponding to $\alpha, \beta$, $\dot\alpha$ and $\dot\beta$ computed for $\gamma=\gamma_j \pm \delta\gamma_j$ and $\dot\gamma = 0$.

To compute the line of variation in $(\theta_x,\theta_y)$ due to non-zero $\dot\gamma$ (binding), we first 
make the assumption that the target orbit is bound, in which case, we can write the system's energy more carefully, and derive (as per \citetalias{Bernstein2000})
\begin{equation}
	\dot\gamma^2 \leq \dot\gamma^2_\mathrm{bind} \equiv 2 GM_\sun \gamma^3 \left(1 + \gamma^2 - 2 \gamma \cos\beta_0 \right)^{-1/2} - \dot\alpha^2 - \dot\beta^2,
\end{equation}
where $\beta_0$ is the target's solar elongation.  The line of variations is then derived from positions for orbits
corresponding to $\alpha, \beta,$ $\dot\alpha$ and $\dot\beta$ computed for $\gamma=\gamma_j$ and $\dot\gamma = \pm \dot\gamma_\mathrm{bind}$.

\begin{figure}[tb]
	\centering
	\includegraphics[width=0.8\textwidth]{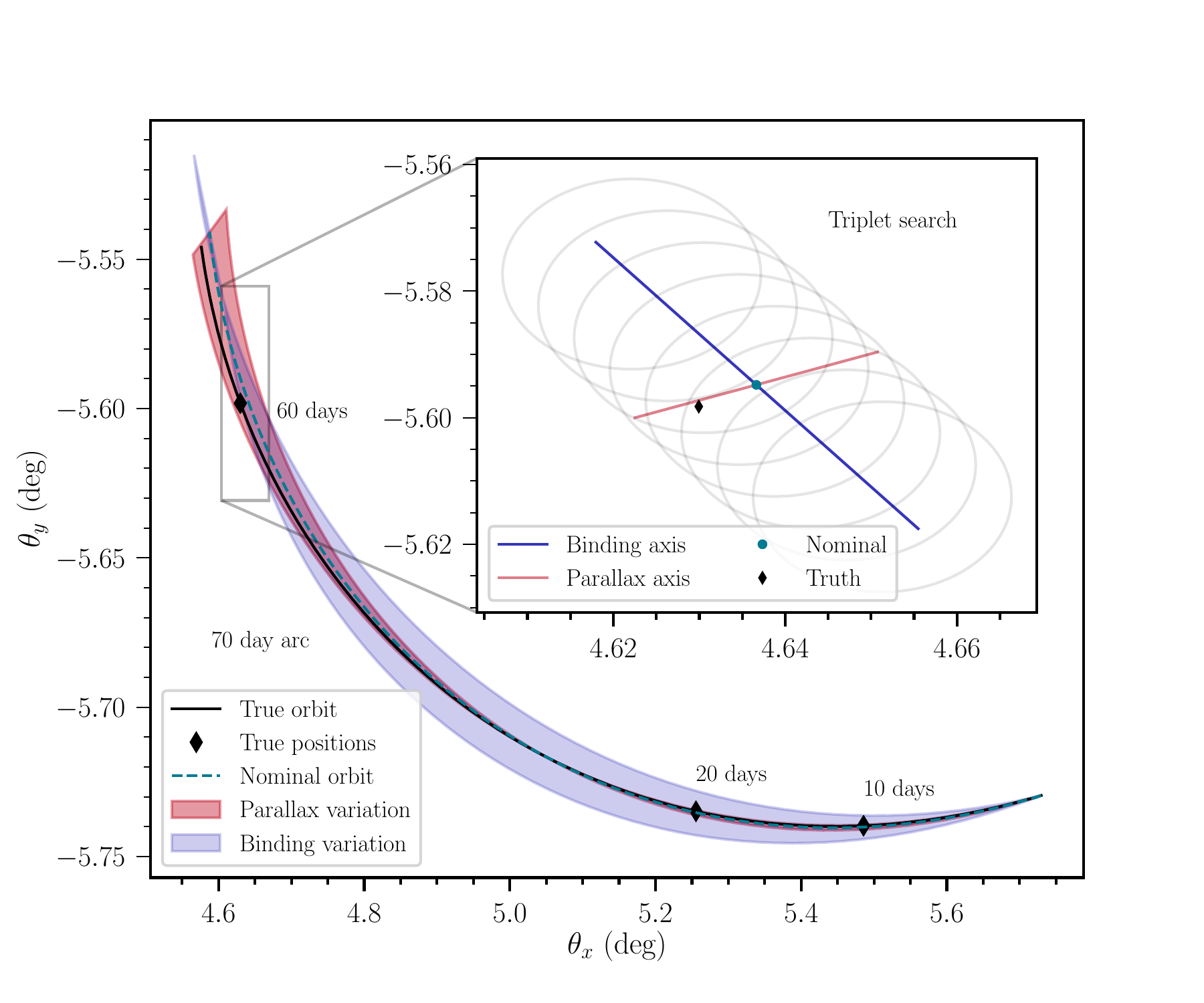}
	\caption{Example of a target orbit at a distance $d = 34 \, \mathrm{AU}$, with detections 10, 20 and 60 days past $t_0$. The search is made with distance bin $\gamma_j  =1/(35 \, \mathrm{AU})$, and $\delta\gamma_j/\gamma_j = 0.05$. The bands show the bounds of predictions vs.\ time of a third position given the pair at 10, 20 days, with the orbit fit to this pair with $\dot\gamma = 0$.  The nominal $\gamma=\gamma_j, \dot\gamma=0$ orbit is the dashed line, while the grey and red ranges show the effects of variations in $\gamma$ (``parallactic'') and in $\dot\gamma$ (``binding''), respectively. The inset shows the position uncertainties at $t=t_0+60$~days. The gray circles show the search region for the third transient, and the diamond is the actual position for this orbit at time $t$.\label{im:triplet}}
\end{figure}

Figure \ref{im:triplet} illustrates how the search region for triplet candidates at some time $t$ is the parallelogram constructed from the parallax axis of $\gamma$ line of variations and the binding axis of $\dot\gamma$ line of variation. 

For each pair, we search for triplets in exposures up to 90 days after the second detection.
For each exposure that crosses the region of potential third
transients, we use its kD tree to rapidly locate for transients lying
in a set of circles that cover the parallelogram of potential orbit extrapolations (plus a small contribution for position measurement errors).

One can show that the leading order behavior in $\gamma$ of the total number of spurious triplets (from randomly positioned transients) is
\begin{equation}
	N_3(\gamma) \simeq n N_2(\gamma) A_3(\gamma) \propto n^3 \gamma^{7/2},\label{eq:n_trip}
\end{equation}
where $A_3$ is the area of the search parallelogram.

The Y4 search generates $6 \times 10^{10}$ triplets (Table \ref{tb:stages}).

\subsection{Growing $n$-lets}
\label{sec:grow}

Once a triplet of detections is identified, we can in principle fit an orbit with all six orbital parameters left free.  We follow routines similar to the ones in \citetalias{Bernstein2000} to fit the orbits and determine expected positions and their linearized uncertainties for the circumstances of any \des\ exposure.  We do not make use of the prior in the \citetalias{Bernstein2000} equations~(20) and (21) that favors nearly-circular orbits, because it can cause convergence problems for some of the very distant and nearly parabolic TNO orbits that we wish to search.  Instead we institute a Gaussian prior on (inverse) distance to force the orbit into our distance bin.  The prior contributes to the fit $\chi^2$ as
\begin{equation}
  \chi^2_{\rm prior} = s_\gamma \frac{(\gamma-\gamma_j)^2}{\delta\gamma_j^2},
  \label{eq:distprior}
\end{equation}
where the strength of this prior is adjusted by setting $s_\gamma$, and we choose $s_\gamma \approx 1.$ We also include a prior for bound orbits, defined as
\begin{equation}
  \chi^2_\mathrm{bind} = b \frac{\dot\alpha^2 + \dot\beta^2+\dot\gamma^2}{2GM_{\sun} \gamma^3}
   = \frac{|\textrm{KE}|}{|\textrm{PE}|}.
	\label{eq:bindprior}
\end{equation}
Here, $b$, the binding factor, defines the strength of this
prior. Note that this is equal to $b$ for a parabolic orbit. We set $b
= 4$ for our fits.  The $\chi^2_{\rm bind}$ is added to the quantity
being minimized in the BK code.

The growing process starts with the orbit fitted to a triplet.  The position and error ellipse are calculated at all other exposures that the TNO might cross. The kD tree for each potential exposure is searched for transients that lie within the $4\sigma$ error ellipse defined by convolving the error ellipse of the orbit prediction at the time of the search exposure with the measurement uncertainty of each transient position on the exposure.  For each such transient found, a new $n$-let is defined, and is queued for its own orbit fit.  The addition of transients is iterated for each $n$-let until no new transients are found to be consistent with the orbit.  We also discard any $n$-let whose orbit fit is significantly unbound, whose best-fit orbit has unacceptably high $\chi^2$ value, or which duplicates a set of transients that have already been examined.

Each time we are attempting to grow an $n$-let, we calculate the \emph{false-positive rate (FPR)} for the addition of the $(n+1)$th transient.  If the $4\sigma$ error ellipse on exposure $j$ has area $A_{j, \rm search},$ and the density of transients on this exposure is $n_{j},$ then the total probability of a spurious linkage is
\begin{equation}
  \mathrm{FPR} = \sum_j A_{j,\rm search} n_{j},
\end{equation}
where the sum is over all exposures being considered for the $(n+1)$th transient.      
Our calculation of FPR currently accounts for CCD gaps and other details of geometry only approximately, but this will be sufficient.
Note that as the orbit becomes better defined from higher $n$ and longer arc, the FPR for further additions will shrink. If an $n$-let has short arc and we are searching for transient $n+1$ in a different season's exposure, then the error ellipse may be large, and we will be flooded with false linkages.  We therefore do not search an exposure if the contribution of this exposure to the FPR sum would be $ A_{j,\rm search} n_j>10.$  Since the transient catalog density is $\lesssim 200\, \textrm{deg}^{-2},$ this is roughly a requirement that the orbit be localized to an area of $\lesssim 0.05 \,\textrm{deg}^2$ if we are to proceed.  We do not implement this FPR cutoff until the $n$-let has linked detections on $\ge4$ distinct nights, since the $A_{j,\rm search}$ is unavoidably large with $\le3$ points on the arc.




The growing process terminates when no additional detections are found to match the orbit fit to an $n$-let.  We associate with this terminal $n$-let the FPR that was calculated for the final \emph{successful} linkage step.  Thus the FPR recorded with the $n$-let estimates the probability that the last detection linked to the orbit is spurious, i.e. a transient that randomly fell within the orbit's error ellipse.  In the future this information will allow us to estimate the number of spurious linkages contaminating our TNO catalog.   At present, we only make a very mild cut of discarding individual $n$-lets with $\mathrm{FPR}>1$.  If a terminal $n$-let has detections in 7 or more unique nights and has $\mathrm{FPR}< 10^{-3}$, we call it a \emph{secure} orbit, and we remove all of its detections from consideration for linkage to further orbits.

\subsection{Reliability of an orbit}\label{sec:reliab}
At the end of the linking process, we have to decide which sets of linked detections reliably correspond to multiple detections of the same real solar system object, vs.\ spurious linkages of mixtures of detections of distinct sources or artifacts.  In this Y4 search, we use the following criteria to cull the set of unique terminal $n$-lets to high-reliability candidates.
\begin{enumerate}
	\item If the number of unique nights on which these detections were seen is $\mathtt{NUNIQUE}<6,$ the $n$-let is rejected.  Of the 132 linkages with $\mathtt{NUNIQUE}=6$ meeting the other criteria, only 52 were confirmed as real by the method of Section~\ref{sec:subthreshold}.
The number of $n$-lets with $\mathtt{NUNIQUE}=5$ is large and certainly dominated by spurious linkages.  We defer until the final \des\ search an effort to extract a reliable candidate set from these. 1684 distinct linkages satisfy this criterion.
    \item The time between first and last detections ($\mathtt{ARC}$) must satisfy $\mathtt{ARC}>6$~months, i.e. we demand detections in distinct \des\ observing seasons.  More stringently, we define $\mathtt{ARCCUT}$ to be the shortest arc that remains after eliminating any single night of detections, and demand also that $\mathtt{ARCCUT}>6$~months.  In effect this means that at least two detections must occur outside the season of the triplet that gestates the $n$-let. 6 linkages have $\mathtt{ARC} < 6 $ months, and another 1235 also have $\mathtt{ARCCUT} < 6$ months.
	\item The $\chi^2$ per degree of freedom from the (prior-less) orbit fit must satisfy $\chi^2/\nu < 4$, which rejects another 19 linkages.
	\item The FPR is less than $1$. No surviving linkages fail this criterion.
\end{enumerate}
A total of 424 distinct linkages pass these criteria in the Y4 search (see Table \ref{tb:stages}).
The ``sub-threshold confirmation'' technique described next will serve to remove spurious linkages from the candidate list defined by the above criteria.

\section{Sub-threshold confirmation}
\label{sec:subthreshold}
We assume without further investigation that all candidates with $\mathtt{NUNIQUE}>10$ are real TNOs, since the calculated FPR's for these are extremely small.
For objects with $6 \leq \mathtt{NUNIQUE} \leq 10$, we test the reality of each candidate TNO by searching for its presence in exposures that did \emph{not} yield a detection of the source, but which the best-fit orbit suggests should contain an image of the object.  The concept here is: if the object is real, it is lurking just below detection threshold in these non-detection images, and by stacking such exposures along the orbit, we will obtain a significant detection.  If, however, the orbit is spurious, then there should be no excess flux in the non-detection images along the (meaningless) orbit.  Given that we typically have $\approx 10$ non-detections (excluding the $Y$ band), the non-detection stack should be $\approx1.25$ mag deeper than a typical single image.

The non-detection images are by definition going to have lower mean $S/N$ ratio on a real TNO than the typical detection image.  This could be because of poorer observing conditions; or being in filters with less favorable $S/N$ given the color of the TNO; or at fainter points on the light curve of a given TNO. Furthermore there are no degrees of freedom in this sub-threshold confirmation significance, because the orbit is fixed to the best fit to the detected transients.
So the appearance of a signal even at $S/N=4$ in the non-detection stack should be considered a strong confirmation.

We proceed by first measuring a flux for the putative TNO in every image $\mu$ (both detections and non-detections)  that contains the best-fit orbit.  For exposure $\mu$ in band $b$, we compute a windowed, sky-subtracted flux $f_{\mu,b}$ as:
\begin{equation}
	f_{\mu,b} = k_\mu\sum_j (s_{\mu,j} - \hat{s}_\mu) W_\mu(\mathbf{x}_j).
\label{stcflux}
\end{equation}
Here, $s_{\mu,j}$ is the photon count at pixel $\mathbf{x}_j$,  and $\hat{s} _\mu$ is the sky background flux, computed by taking the mean flux in an annulus centered at the nominal position and with inner radius $8\arcsec$ and outer radius $10\arcsec.$  For the photometry window $W_\mu(\mathbf{x}_j)$ we adopt a circular Gaussian centered on the position predicted from the orbit, and having a FWHM of 1\arcsec.  This window will retrieve near-optimal $S/N$ ratio for the putative TNO in typical \des\ seeing.  The factors $k_\mu$ remove the variations in the photometric zeropoints of the exposures, placing fluxes on a common scale \citep{Burke2017}.   The RMS 
noise $\sigma_{\mu,b}$ in $f_{\mu,b}$ is calculated by propagating the Poisson noise of the $s_{\mu,j}$ through Equation~(\ref{stcflux}).

We need to remove the contribution of static sources from $f_{\mu,b}.$ To do so we apply (\ref{stcflux}) to the coadd image at this location of sky to obtain 
flux $f_{\mu,b,\mathrm{coadd}}$ and noise $\sigma_{\mu,b,\mathrm{coadd}}$.  If the static flux is a significant fraction of the inferred flux, we discard this exposure from the calculation.  The criteria for rejection are
$f_{\mu,b} < 3 f_{\mu,b,\mathrm{coadd}}$ (the TNO flux does not dominate the static flux) and $f_{\mu,b,\mathrm{coadd}} > 5 \sigma_{\mu,b,\mathrm{coadd}}$ (nonzero static flux is confidently detected). 
Single exposures failing this cut are ignored in further evaluation of their TNO orbit.

The inverse variance weighted total flux, then, coming from the remaining images is
\begin{equation}
	\hat{f}_b = \dfrac{\sum_\mu f_{\mu,b}\sigma_{\mu,b}^{-2}}{\sum_\mu \sigma_{\mu,b}^{-2}},
\end{equation} 
with variance
\begin{equation}
	\operatorname{Var}(\hat{f}_b) = \frac{1}{\sum_\mu \sigma_{\mu,b}^{-2}}.
\end{equation}

The \emph{significance} of a detection in band $b$, then, is
\begin{equation}
	\mathcal{S}_b = \frac{\hat{f}_b}{\sqrt{\operatorname{Var}(\hat{f}_b)}}.\label{eq:signif}
\end{equation}

For visual inspection, we also combine the individual exposures' images of the putative TNO using inverse-variance weighting:
\begin{equation}
	\mathbf{Img}_b = \frac{\sum_\mu k_\mu\mathbf{Img}_{\mu,b} \sigma_{\mu,b}^{-2}}{\sum_\mu \sigma_{\mu,b}^{-2}},\label{eq:summed}
\end{equation}
where $\mathbf{Img}_{\mu,b}$ is a $50 \times 50$ pixel image cutout centered on the detection at exposure $\mu$.

To compute the \emph{total significance}, we combine all detections in the $griz$ bands, by first transforming all $giz$ fluxes into an $r$ band flux, assuming nominal colors corresponding to the bimodality break for Centaurs : $g - r = 0.75$, $r - i = 0.25$ and $r - z = 0.5$ \citep{Ofek2012,Pike2017}. To confirm an orbit, we only use the images in which there is no detection in the Final Cut catalog.  We define the ``sub-threshold significance'' (\texttt{STS}) to be the value of $\mathcal{S}_b$ from Equation~(\ref{eq:signif}) evaluated only on these non-detection images.
Spuriously linked detections cannot contribute to this total significance. Non-detection images taken within 1 hour of a detection are omitted from this summation, so that a spuriously linked asteroid (or image defect) cannot recur in a non-detection image and contaminate
the \texttt{STS}.  In other words, we want the \texttt{STS} to be statistically independent of the detections in the absence of a true TNO.

\begin{figure}[htb]
	\centering
	\includegraphics[width=\textwidth]{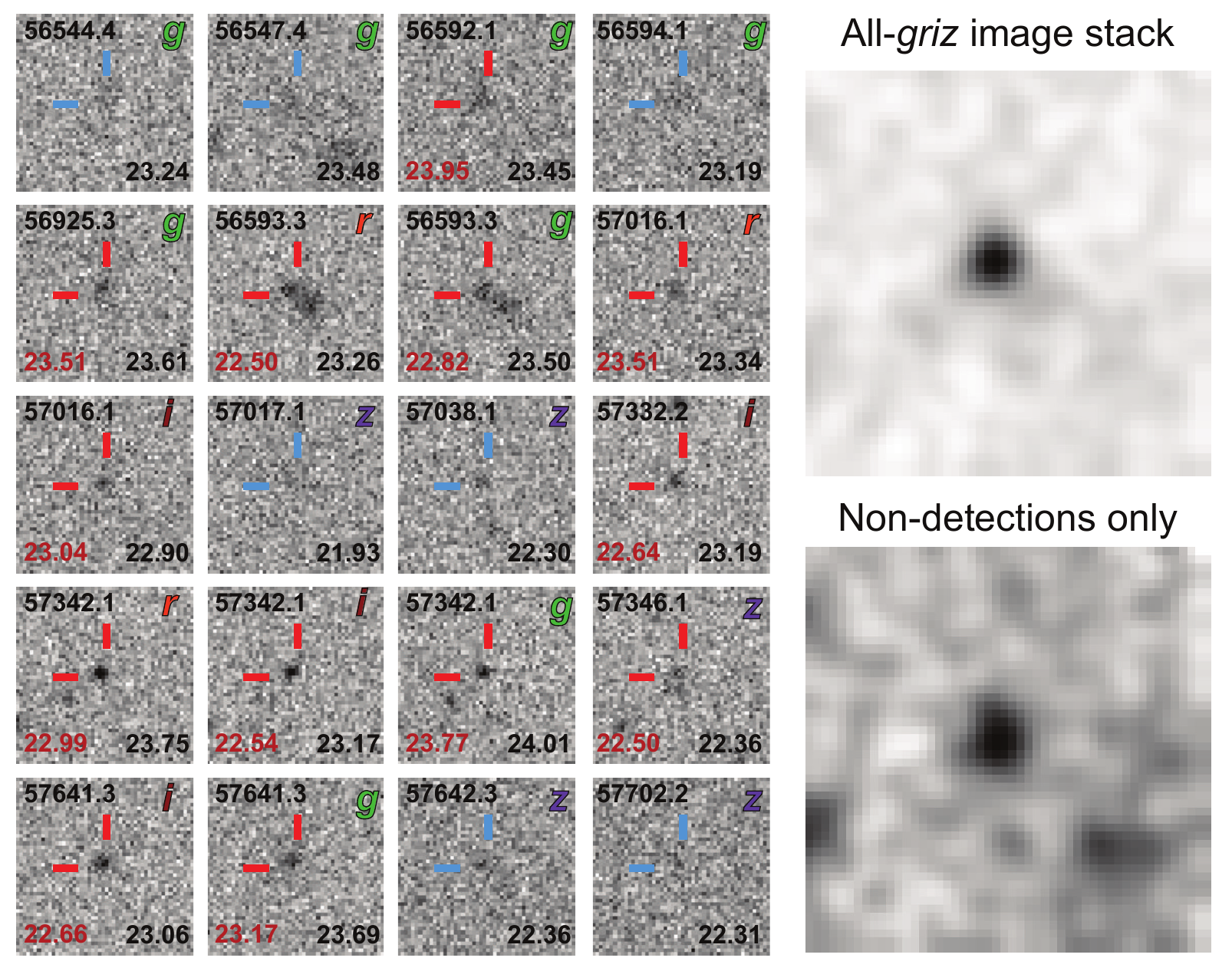}
	\caption{\emph{Left:} Postage stamps of the detected and expected positions for one of the recovered TNOs with $\mathtt{NUNIQUE}  = 8$. The top of each stamp shows the MJD when the exposure was taken and its band, with $m_{50}$ for the exposure at lower right. Exposures where the object was detected are marked in red crosshairs and have the measured magnitude listed on the lower left. Exposures of the putative TNO which did not yield a detection are marked with blue crosshairs. \emph{Right:} Summed $griz$ images (all images: top; non-detections only: bottom) convolved by a $1\arcsec$ Gaussian kernel.  The lower-right image shows a highly significant detection $(\mathtt{STS}=12.93)$, and is considered confirmed.\label{im:snapshot_real}}
\end{figure}

\begin{figure}[htb]
	\centering
	\includegraphics[width=\textwidth]{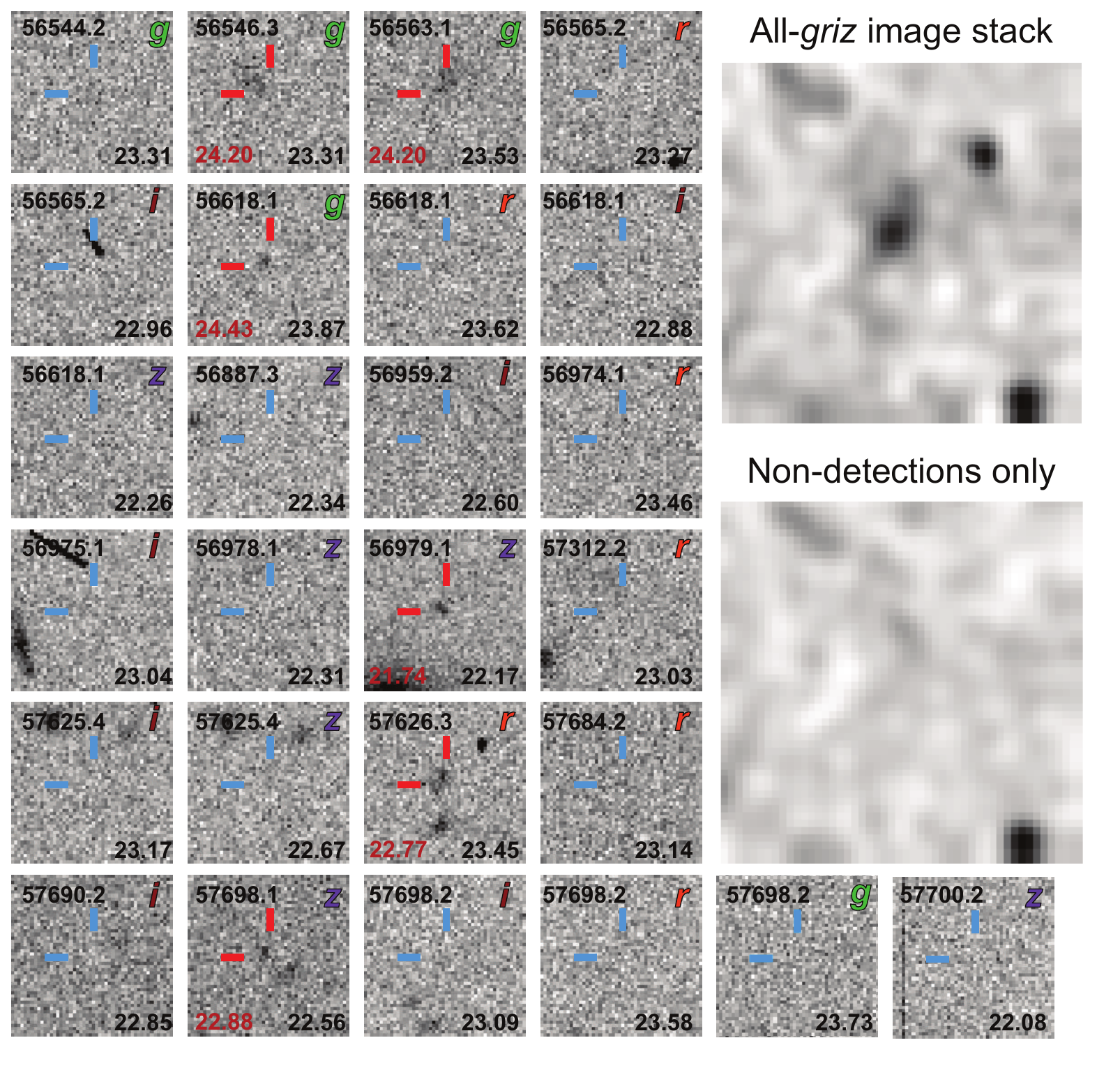}
	\caption{Similar to Figure \ref{im:snapshot_real}, but showing a candidate that does not pass the sub-threshold significance test. The $\mathtt{STS}$ of this candidate is $-0.16$. \label{im:snapshot_false}}
\end{figure}

\begin{figure}[htb]
	\centering
	\includegraphics[width=0.59\textwidth]{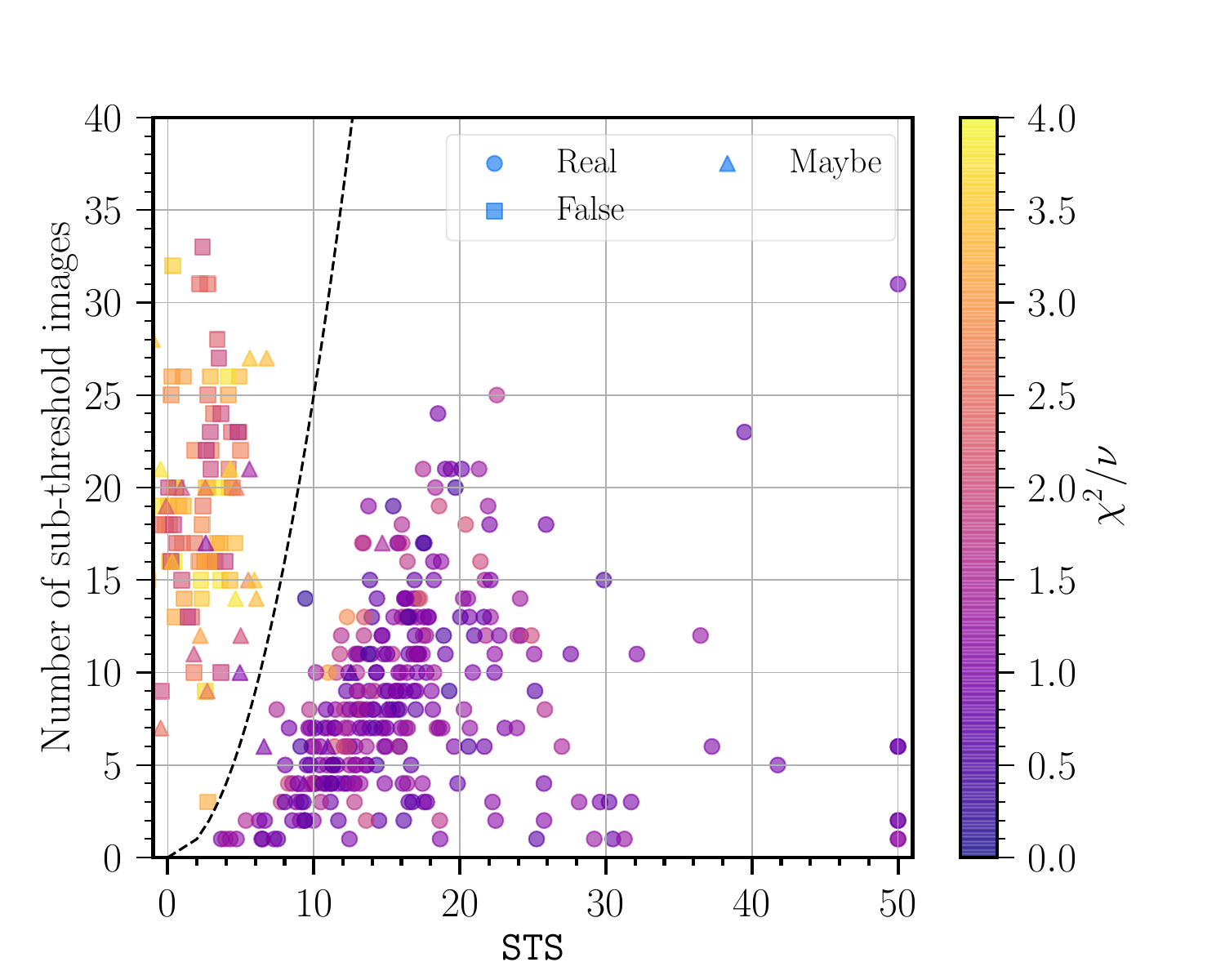}
	\caption{$\mathtt{STS}$ versus number of images used in the sub-threshold sum, as well as a color code for $\chi^2/\nu$.  The symbol shapes encode the $\mathtt{R}$, $\mathtt{F}$, $\mathtt{M}$ status derived from visual inspection. The dashed curve is the $\mathtt{STS} = 2 \sqrt{N_\mathrm{images}}$ parabola, which cleanly divides all ``Real'' from ``False'' linkages and separates the two obvious groupings in the plot.  We consider all 316 linkages lying to the right of the parabola as being confirmed TNOs.\label{im:sts}}
\end{figure}

Figure \ref{im:snapshot_real} shows an example of a real TNO that passes this test. Figure \ref{im:snapshot_false} shows a spurious linkage.

 In addition to the value of $\mathtt{STS}$, we also vet all 424 candidates passing the basic parametric cuts of Section~\ref{sec:reliab}
by visually inspecting postage stamps of all images of the putative orbit.  We also view the summed $griz$ images (following Equation \ref{eq:summed} and using the nominal colors) created both using all the images, and the image created from just the non-detection images. Each candidate received a score of \texttt{R} (real), \texttt{F} (false) or \texttt{M} (maybe, corresponding to an unsure classification) independently by three of the authors (PB, GB, and MS). In every case, at least two graders agree, and we accept the majority classification.
The visual inspection also identifies non-detection images in which the \texttt{STS} calculation is being contaminated by any static sky object or unmasked defect that sneaked past the cuts on coadd flux described above.  For these cases we recalculate the \texttt{STS} after purging this defective exposure.  Note that the grading was done while blind to the orbital elements of the detection.

Figure \ref{im:sts} shows that the 424 orbit candidates fall into two distinct groups on the plane of \texttt{STS} vs.\ $N_{\rm images},$ the number of non-detection images contributing to \texttt{STS}.  These groups are cleanly separated by the parabola $\mathtt{STS} = 2 \sqrt{N_\text{images}},$ and every candidate graded as \texttt{R} (\texttt{F}) falls to higher (lower) \texttt{STS} than this curve, suggesting that cutting objects to the left of this curve is a very efficient means of removing spurious linkages. Our confidence in this cut is boosted by noting that the linkages failing this cut are generally much worse fits to an orbit than the retained sample.  As illustrated by Figure~\ref{im:chi2dist}, however, the $\chi^2/\nu$ statistic is not by itself a sure-fire discriminant between real and spurious linkages.  We might indeed expect that some spurious linkages attain low $\chi^2$ by luck, and also that some true TNOs might have inflated $\chi^2$ values because they are binaries with significant photocenter motion about the center of mass.  So we do not impose any stricter cut on $\chi^2/\nu.$

A total of 316 linkages pass the \texttt{STS} cut, and comprise the complete sample of detections from the Y4A1 \des\ wide-survey search see Table \ref{tb:stages}).

The weighting of the different bands in the \texttt{STS} value yields optimum (high) values for objects with mean TNO colors.  For individual objects with colors deviating from this mean, the \texttt{STS} we compute is non-optimal (lowered). So the technique, applied with fixed nominal weighting, cannot produce false positives, but might produce a false negative if an object's colors are significantly different from the ones chosen.
The calculation is, however, robust: object 2013 $\mathrm{SN}_{102}$ is one of our reddest at $g - r \approx 1.23$ (compared to the nominal $g - r = 0.75$; the $r - i$ and $r - z$ colors of the object are within 20 mmag of the nominal colors). Its \texttt{STS} is $\approx 22.7$ with nominal weighting and rises to $\approx 24.1$ by computing this value with weights optimized to its colors. Thus the nominal weighting yields \texttt{STS} only 6\% lower than optimal in this extreme case.

The \texttt{STS} test can be also be applied to objects that are too faint to be recovered in the survey, but whose orbits are known to good precision. As an example, we have measured the \texttt{STS} of 2016 $\mathrm{QU}_{89}$, discovered on the deep \des\ fields \citep{Khain2018} and too faint to be detected in the wide survey. This object was recovered in only two of the images we analysed here, despite being inside a functional CCD in 17 images. We have measured its \texttt{STS} to be $24.6$, being well above the required $ 2 \sqrt{N_\mathrm{images}}\sim 8.2$ required to recover it.

Finally, we note that the tools presented here allow us to distinguish real from spurious 5-night detections by repeating the $\mathtt{STS}$ test, although this will require us to evaluate many spurious linkages.  The entire linkage process can also be extended to distances closer than 30 AU, but at a significant increase in computational expense (see Equations \ref{paircount} and \ref{eq:n_trip}), perhaps requiring a trade-off between the searched distances (i.e. $\gamma$) and the maximum time span of the pairs and triplets ($\Delta t$).

\section{Completeness testing with synthetic TNOs}
\label{sec:fakes}

In order to test our linking efficiency, we insert a synthetic population of TNOs (``fakes'') into our transient catalog that simulate the observational properties a real object would have.  This is not an end-to-end simulation: the fakes are inserted into the catalogs, not the images.  The cost of doing the latter and reprocessing all the images would be prohibitive.  This shortcut is, however, acceptable, since the
point-source detection efficiencies described in Section~\ref{sec:m50} allow us to assess the detectability of a TNO of given magnitude in a given exposure.  This method does not, however, account for the inefficiency in transient detection that arises
when TNOs fall on or near stationary sources (such that they are part of blended detections, or flagged as static sources), or when TNOs lie on  CCD defects or other artifacts (such that they are flagged as defective and ignored).  A few percent of the active imaging area is lost to such effects in typical exposures.  Image-injection experiments under way on the completed \des\ imagery will quantify this loss and be incorporated into the detection simulator that will accompany the final \des\ TNO catalog \citep{Balrog}. 

\subsection{Simulating the \des\ observations}
The process for simulating the transient-catalog entries for a TNO of given orbital parameters and magnitude $m$ is as follows:
\begin{enumerate}[label=\roman*.]
	\item We find all exposures for which the DECam field of view contains the TNO position given its orbit.
	\item An observed position for the TNO is derived by adding observational error to the position predicted by the orbit.  Currently the observational error is drawn from the distribution of errors for all point sources in the image (including both shot noise and the atmospheric turbulence contribution per Equation \ref{astsigma}); in the future we will properly track the errors vs.\ source magnitude. 
	\item We check whether the ``observed'' position lies within a functional DECam CCD, i.e. we would have collected an image of this TNO. If this is true, this is considered an \emph{observation};
	\item We compute the value of $p(m)$ from Equation \ref{eq:completeness} with the fitted parameters for this exposure and compare to a random value $r$ between 0 and 1. If $p(m) > r$, this observation is considered a \emph{detection} and is entered into the transient catalog. 
  	\item For completeness, a random magnitude error is drawn from the magnitude error distribution determined for this exposure and added to the truth magnitude, for each synthetic transient. Our linking algorithm makes no use of the detected magnitude, so this is irrelevant.
\end{enumerate}

This process does not simulate the loss of TNOs due to potential overlap with other sources or image artifacts, which as noted above creates a loss of a few percent sources.  Note also that no light-curve variation is placed on the simulated sources.

\subsection{Fake population inserted into the transient catalog}
\label{sec:fakes_cat}
We generate a population of synthetic TNOs that is intended to sample the full phase space of possibly detectable orbits and magnitudes, with no intention of mimicking the true TNO population.
We generate the fakes by sampling the barycentric phase space $\{ x_0, y_0, z_0,\dot x_0, \dot y_0, \dot z_0\}$ at a reference time $t_0$ near the survey midpoint. To generate the position vector, we sample the unit sphere by constructing uniformly distributed angles with a Fibonacci lattice \citep[see, \emph{e.g.},][]{Gonzalez2009} in equatorial coordinates, and discard all points well outside the \des\ footprint. Each fake is assigned a random barycentric distance from a distribution placing half of the fakes uniformly in the range 30--60~AU and half  logarithmically distributed between $60 \AU$ and $2500 \AU$. 
Similarly, the velocities are sampled by placing angles on a spherical Fibonacci lattice, and assigning a velocity $v=f^{1/3}v_{\rm esc}(d),$ where $v_{\rm esc}(d)$ is the escape velocity at the barycentric distance, and $f$ is a uniform deviate between 0 and 1.  

Each fake is also assigned an $r$ band magnitude, independent of its distance, sampled from a uniform distribution between $m_\mathrm{bright}=20$ and $m_\mathrm{faint}=24.5$, spanning the range in which we expect our Y4 discovery efficiency to go from near unity to zero. Each TNO's magnitude is assumed to be constant (\emph{i.e.} there is no lightcurve for the object). The colors are fixed and chosen to be similar (but not identical) to those of (136199) Eris \citep{Eris} as observed by \des: $g -r = 0.55$, $r - i = 0.07$,  $r - z = -0.02$ and $r - Y = -0.04$. 

The transient catalog used in the search includes these fakes, and the process is blinded in that the linking algorithms make no distinction between real and fake transients. 

\subsection{Completeness estimates}

\begin{figure}[htb]
	\centering
	\includegraphics[width=0.48\textwidth]{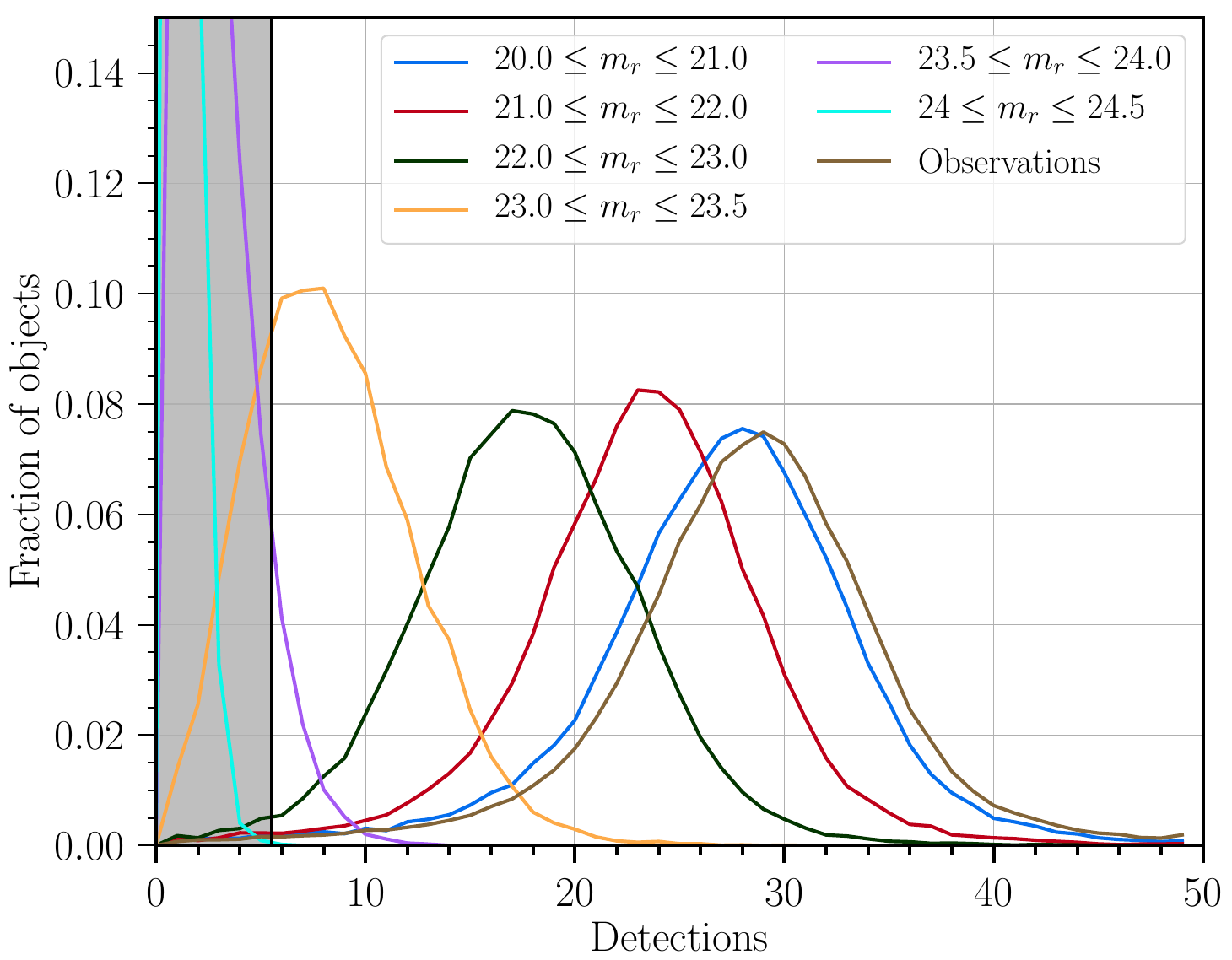}
	\includegraphics[width=0.48\textwidth]{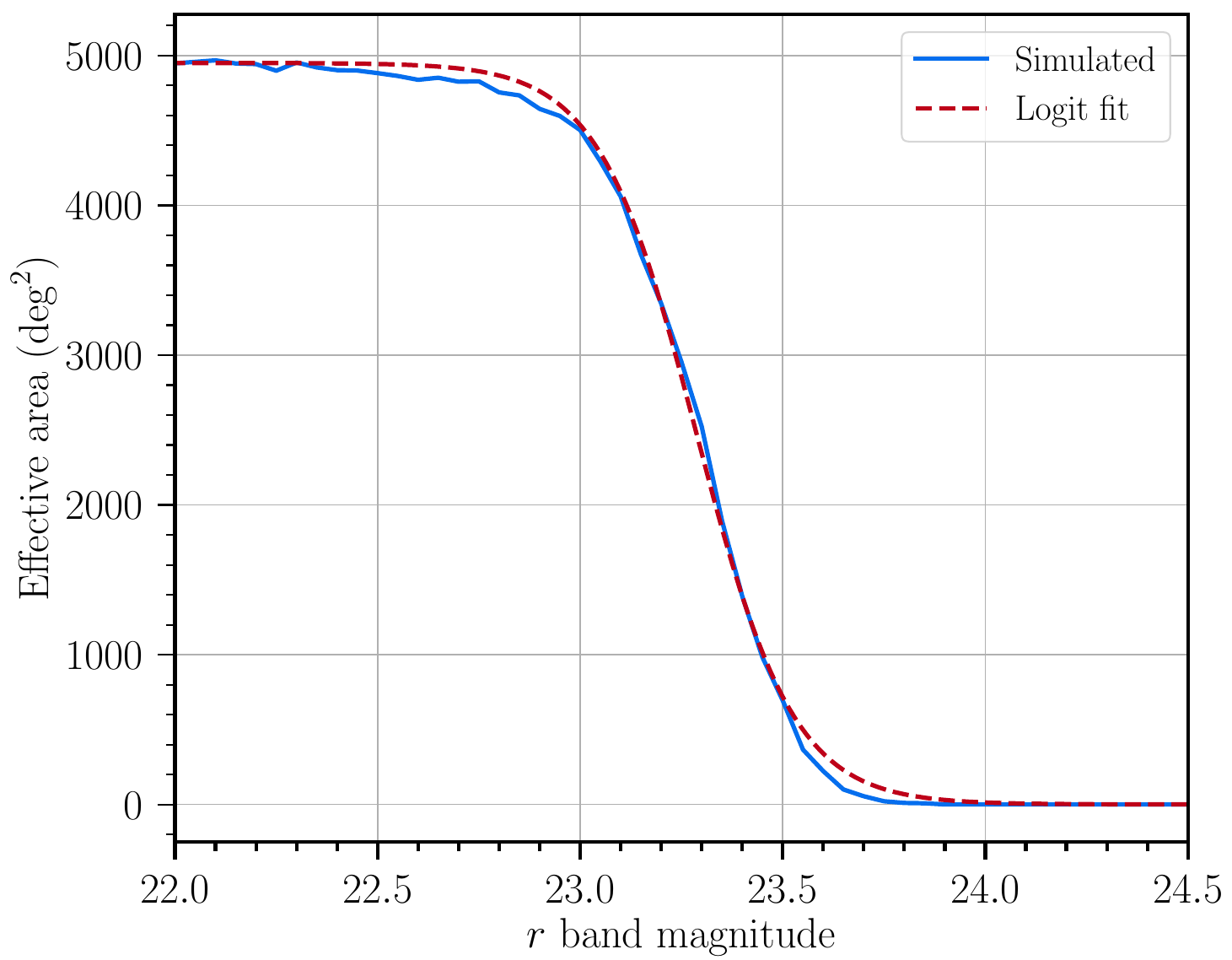}
	\caption{\textit{Left:} The brown curve plots the distribution of the number of \des\ observations 
          (i.e. the TNO falls on a CCD) for a sample of 196,663
          simulated TNOs evenly distributed in the \des\ footprint.
          The other curves at left show the distribution of number of
          these observations that would result in detections
          (i.e. signal above detection threshold), as a function of $r$ band magnitude The shaded area indicates the region with less than 6 detections---there are many spurious linkages with $<6$ detections, so we do not yet report TNO detections in this regime. 
         \textit{Right:} Effective search area for an object vs.\ magnitude, averaged over orbital parameters for the simulated TNO population. An object is considered as recoverable if it is detected in more than 6 unique nights. A logit fit (similar to equation \ref{eq:completeness}) to this function shows that $m_{50} = 23.29.$ 
\label{im:fake_detections}}
\end{figure}

\begin{figure}[htb!]
	\centering
	\includegraphics[width=0.49\textwidth]{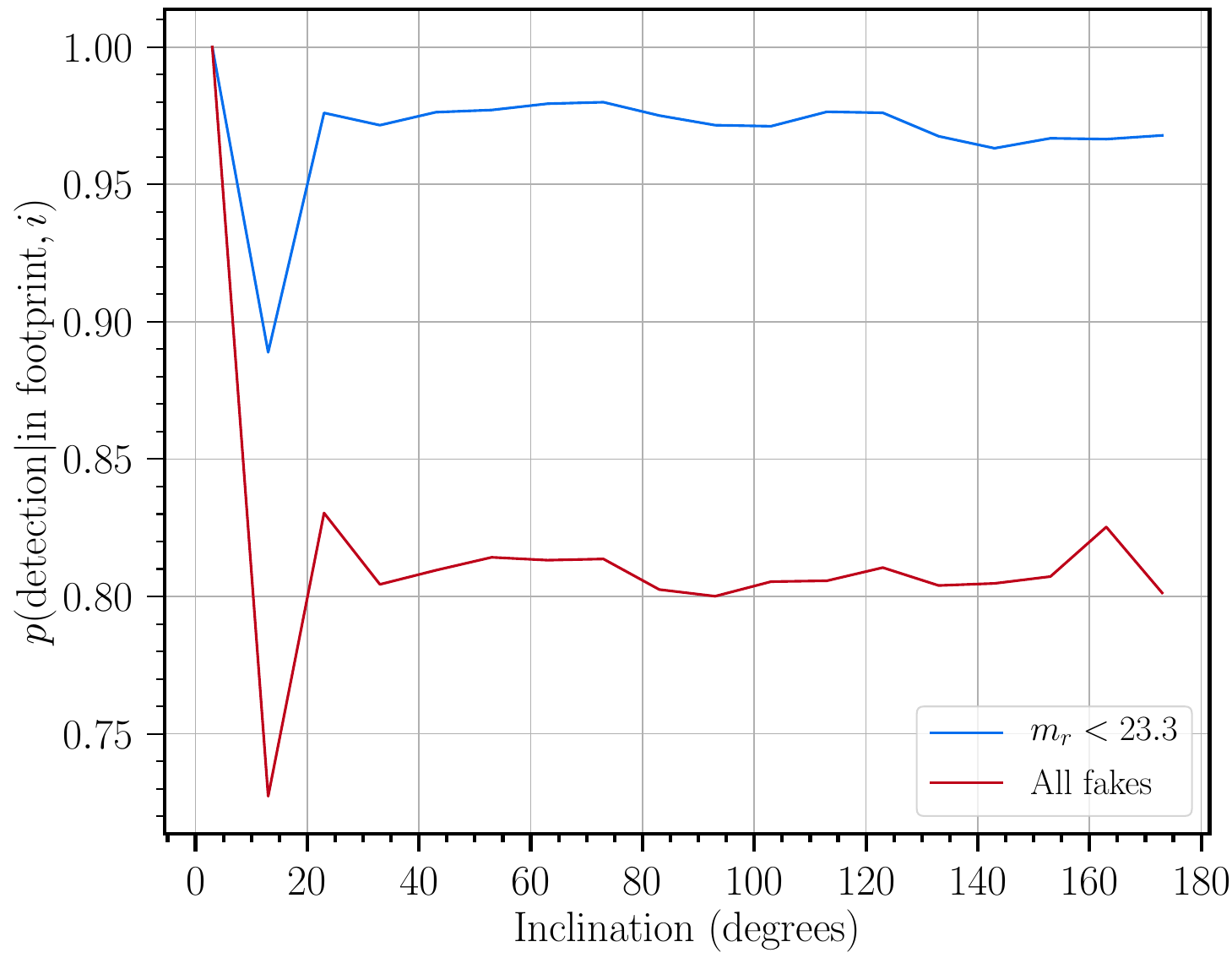}
	\includegraphics[width=0.49\textwidth]{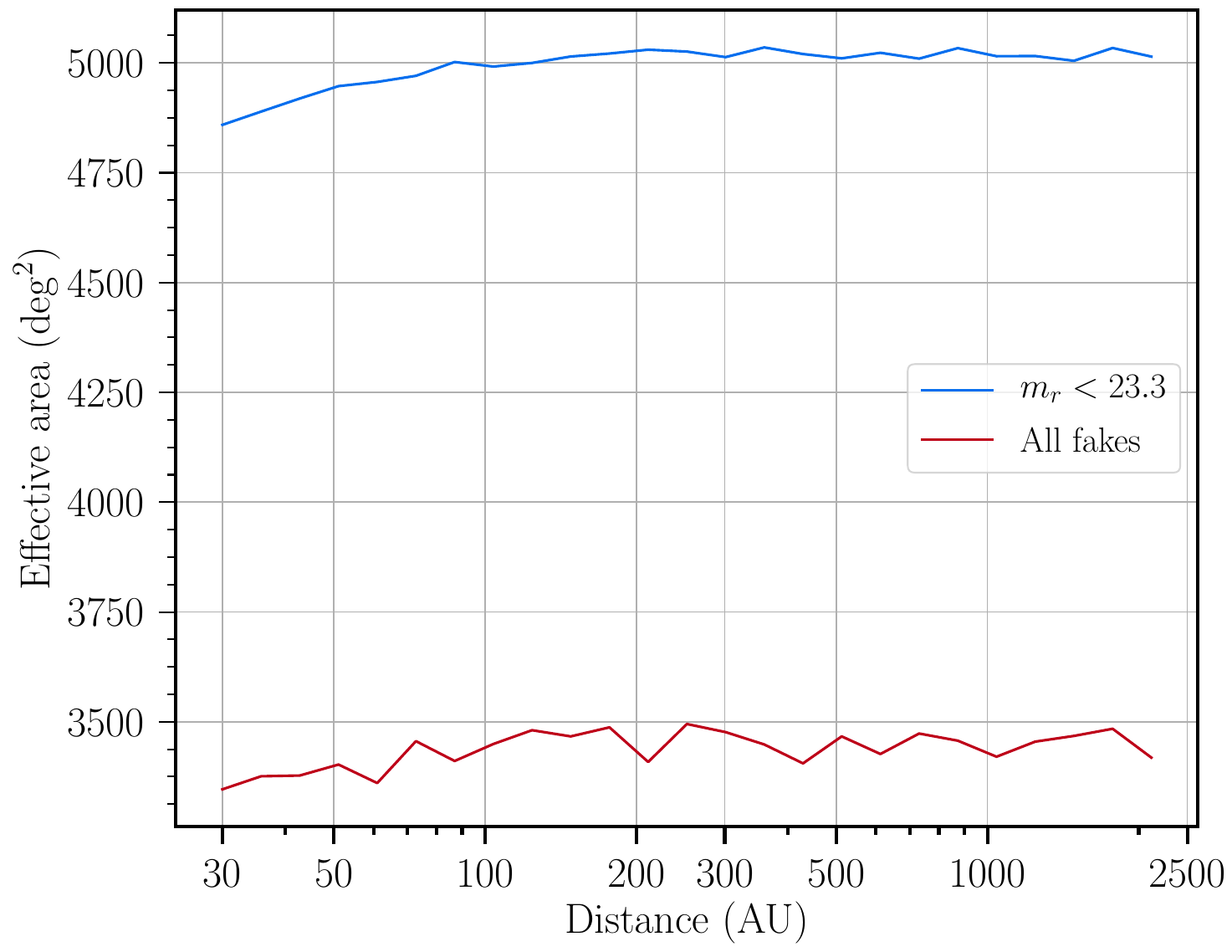}
	\caption{\emph{Left:} Probability of recovery of an object with a given inclination on the complete regime (blue, $m_r < 23.4$) and for all simulated objects (red). Since the survey's longitudinal coverage at low ecliptic latitudes is not as extensive as at high latitudes (see Figure \ref{im:transient_map}), objects with $i \lesssim 10$ and $i \gtrsim 170\degr$ have a higher chance of leaving the nominal footprint. \emph{Right:} Effective search area as a function of barycentric distance, showing very little dependence of recovery probability on an object's distance.\label{im:recovery}}
\end{figure}

Figure~\ref{im:fake_detections}, shows (at left)  the frequency of observations and detections for a population of $\approx200,000$ fakes.
Here we can see that the typical TNO is observed on $\approx20$ distinct nights of the Y4 data, with a tail to low values for TNOs that move in or out of the survey footprint during the survey.  This will occur near the edges and along the thin equatorial stripe of the footprint. Fainter objects are detected in fewer observations, once $m_r>21.$  In the range $23<m_r<23.5,$ about half of fake TNOs are ``recoverable'' by our criterion that they have $\mathtt{NUNIQUE}\ge6.$
The right-hand panel integrates over all orbital parameters to give the effective survey area as a function of $m_r$, indicating that the detection and linkage are highly complete for $m_r<23,$ and 50\% complete at $m_r\approx23.3$ (assuming that all colors are Eris-like). 
Figures \ref{im:recovery} shows the recovery efficiency as a function of inclination and barycentric distance, indicating that there is very little dependence on the orbital properties of an object at fixed apparent magnitude.   The \des\ footprint is much broader than a typical TNO's orbital path, except for the narrow equatorial stripe, meaning that the detectability of a TNO is almost entirely a function of apparent magnitude once it has distance $>30$~AU.  A minor exception is for TNOs with inclination near 0\arcdeg\ or 180\arcdeg, for which a significant fraction of our coverage is in the narrow strip.

\subsection{Completeness over previously known TNOs}
A search of the Minor Planet Center database for known TNOs that were
within the FOV of at least 6 of the \des\ Y4 $griz$ exposures shows
that all such objects above our estimated $r=23.3$~mag 50\%
completeness level are indeed among our 316 detections.  The brightest
of these known TNOs that we do \emph{not} re-discover is
2013~RY$_{108}$, at $r=23.47,$ which was in fact discovered from the
deep supernova-search images in \des.  It is in the FOV for 13
wide-survey exposures on
9 distinct nights, but was only detected in 3 of these exposures.
This is fully consistent with our estimated completeness thresholds.
The final DES TNO search will have a lower single-epoch detection
threshold, and more exposures, that should enable discovery of many
similar TNOs.

\section{Catalog of \des\ TNOs}
\label{sec:catalog}

In Table \ref{tb:catalog}, we present the 316 objects that pass the \texttt{STS} test shown in Figure \ref{im:sts}. 
Additionally, distinct objects found in other searches of \des\ data and reported to the MPC are listed in Table \ref{tb:umich} \citep{Khain2019}. The other searches include data from the deeper, high-cadence supernova fields, and also discovered objects at distances $<30$~AU, which would be missed by the Y4 search.  Since the other searches are not as homogeneous across the \des\ footprint as the Y4 search, the statistical summaries presented here include only objects detected in the Y4 search.

\begin{figure}[tb]
	\centering
	\includegraphics[width=0.49\textwidth]{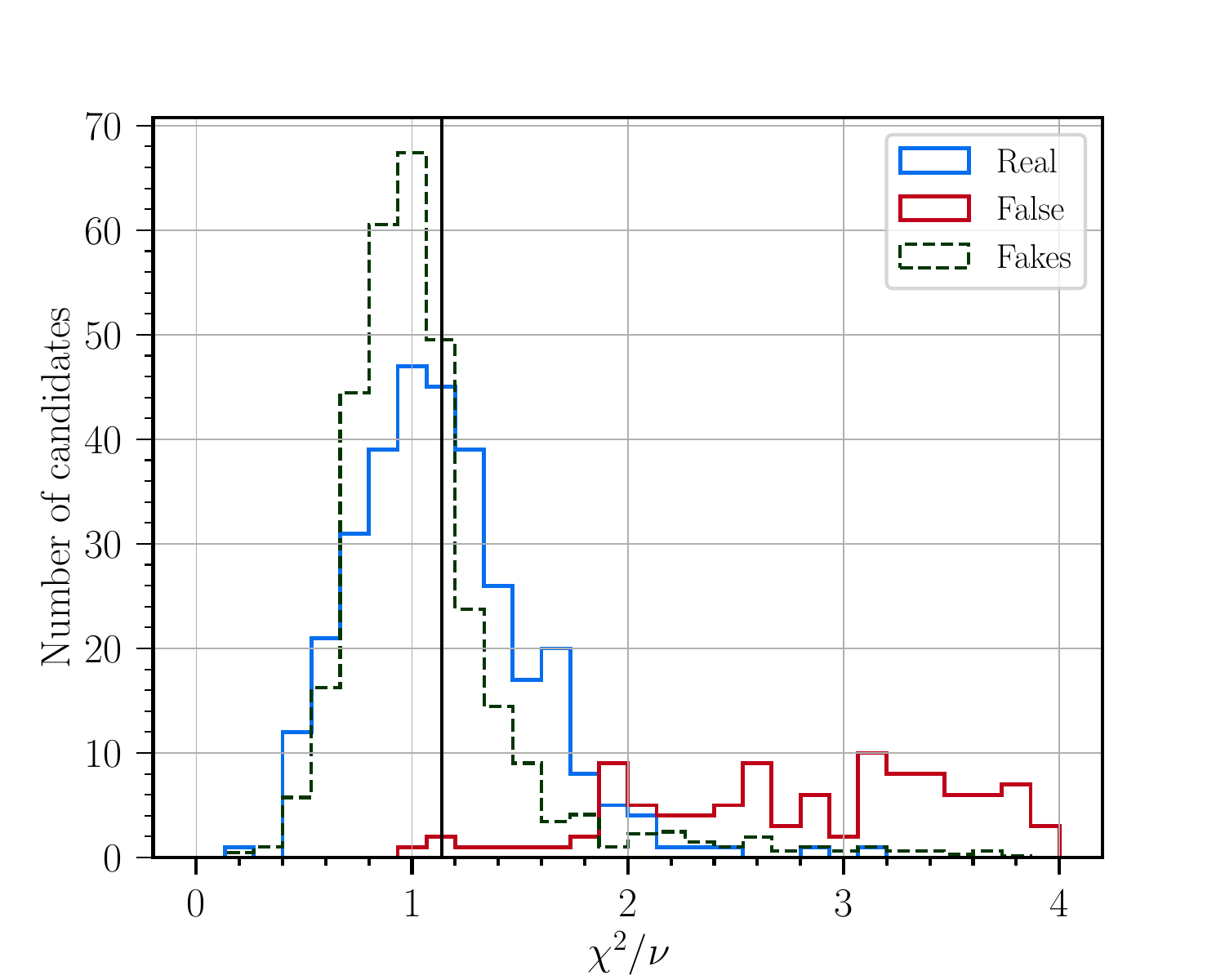}
	\caption{$\chi^2$ per degree of freedom $\nu$ for all 424 candidate orbits, with the blue curve representing the 316 confirmed linkages, and the red representing the candidate orbits rejected as spurious linkages. The black line marks the mean of the blue histogram, $\langle{\chi^2/\nu}\rangle = 1.13$. The dashed curve represents the $\chi^2/\nu$ distribution for the 1727 fakes inserted into the catalog and retrieved by the linking process. The ``Fakes'' histogram is normalized to the same sum (316) as the real detections for easy comparison, showing the real sources to be only slightly poorer fits to their orbits, on average. \label{im:chi2dist}}
\end{figure}

The histogram of the astrometric $\chi^2$ per degree of freedom of best-fit orbits is presented in Figure~\ref{im:chi2dist}. This histogram suggests that our positional accuracies are estimated to good precision, since the peak is close to $\chi^2/\nu=1$.
More precisely, the mean $\langle\chi^2/\nu\rangle=1.13$ suggests that our errors are slightly underestimated.  A similar conclusion can be drawn from comparing to the $\chi^2/\nu$ distribution of the implanted fake detections, for which we know the observational errors exactly, and which yield a slightly lower distribution.
An 11\% increase in the astrometric covariance matrices, corresponding to a $5\%$ increase in errors on positions and orbital elements in real detections, leads to good agreement between the fitted $\chi^2$ and a true $\chi^2$ distribution.  This potential 5\% underestimate of orbital errors should be considered a maximum value, since some of the inflated $\chi^2$ values could instead arise from small photocenter motions in binary TNOs.  A significant fraction of cold classical TNOs ($i \lesssim 5 \degr$) are expected to be binaries \citep{Stephens2006}, and we would not want to cut these from our sample.  The only one of our detected TNOs in W. Grundy's list of known binaries\footnote{\url{http://www2.lowell.edu/~grundy/tnbs/status.html}, accessed Jan. 8, 2020.} is Eris, a large-mass-ratio binary for which photocenter motion should be small, so we cannot yet verify any cases of binary-inflated $\chi^2$.  We note that $\langle \chi^2/\nu \rangle$ for the cold classicals is higher than average at $1.31,$ but we have not yet investigated whether this is attributable to binaries.

\begin{figure}[tb]
	\centering
	\includegraphics[width=0.49\textwidth]{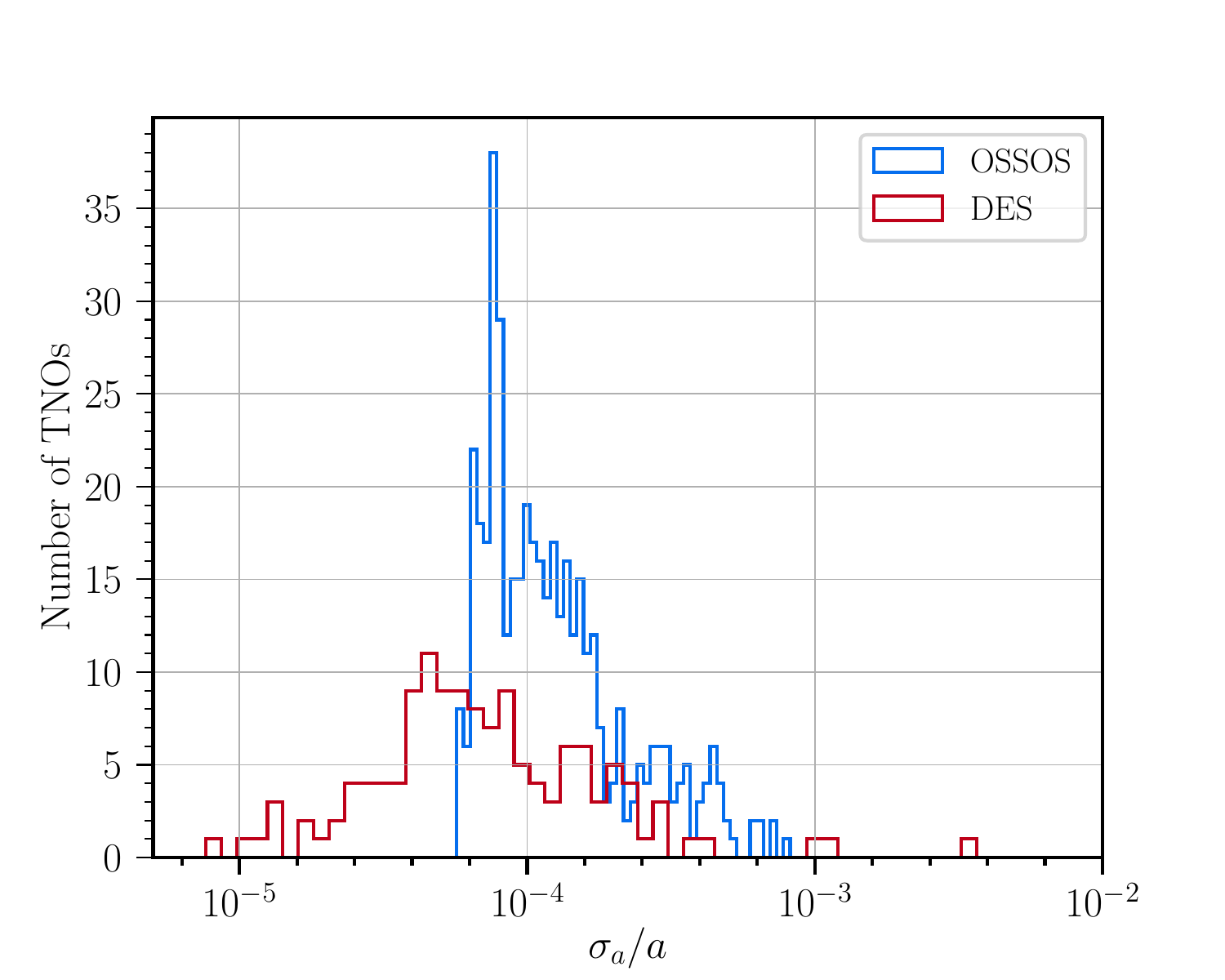}
	\caption{Comparison of $\sigma_a/a$ for the 436 OSSOS \citep{Bannister2018} classical TNOs to the 134 from \des. The median \des\ orbital uncertainty is $\approx1.5\times$ lower than than those from OSSOS, without the need for targeted followup observations.\label{im:ossos}}
\end{figure}

 Figure \ref{im:ossos} compares the quality of the orbits obtained here to the ones from OSSOS \citep{Bannister2018}, where we see that the mean error of the \des\ detections' semi-major axis is lower than those of OSSOS, without the need for targeted followup.  The median \des\ error on $a$ of a classical KBO is $\approx1.5$ lower than in OSSOS, and  will decrease with inclusion of the final 2 years' survey data.

\begin{figure}[tb]
	\centering
	\includegraphics[width=0.99\textwidth]{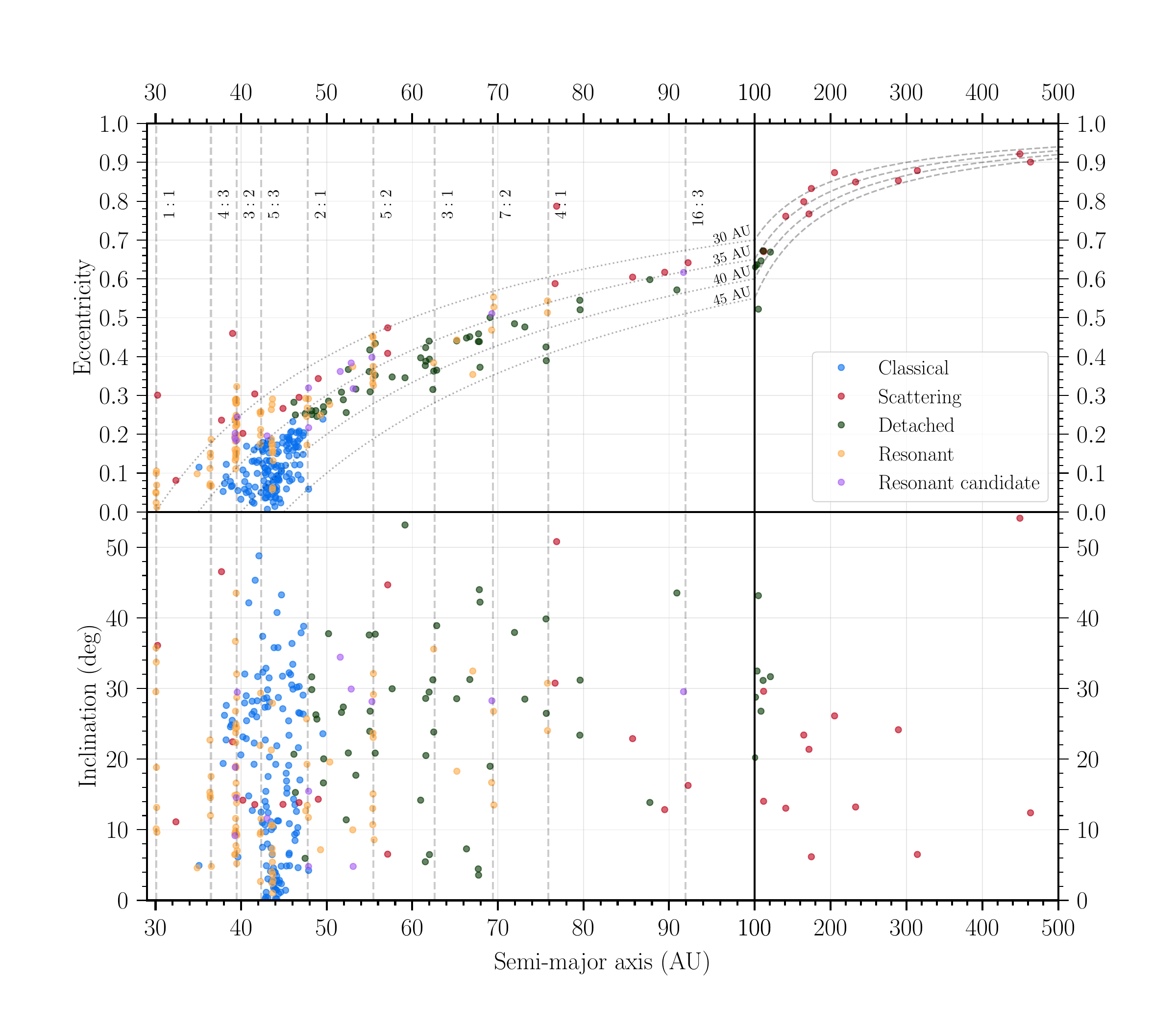}
	\caption{Semi-major axis versus eccentricity and inclination for the 316 TNOs found in \des\ data. The approximate semi-major axis location of some resonances with Neptune is shown by the dashed vertical lines, and the dotted lines in the upper panel represent lines of constant perihelion, color-coded by which dynamical population each object belongs to. Table \ref{tb:pop} lists the number of objects per dynamical class and resonance. We note here that some objects near high order mean motion resonances with Neptune might be identified as ``detached'' due to uncertainties in the orbit parameters.\label{im:aei}}
\end{figure}

\begin{figure}[tb]
	\centering
	\includegraphics[width=0.49\textwidth]{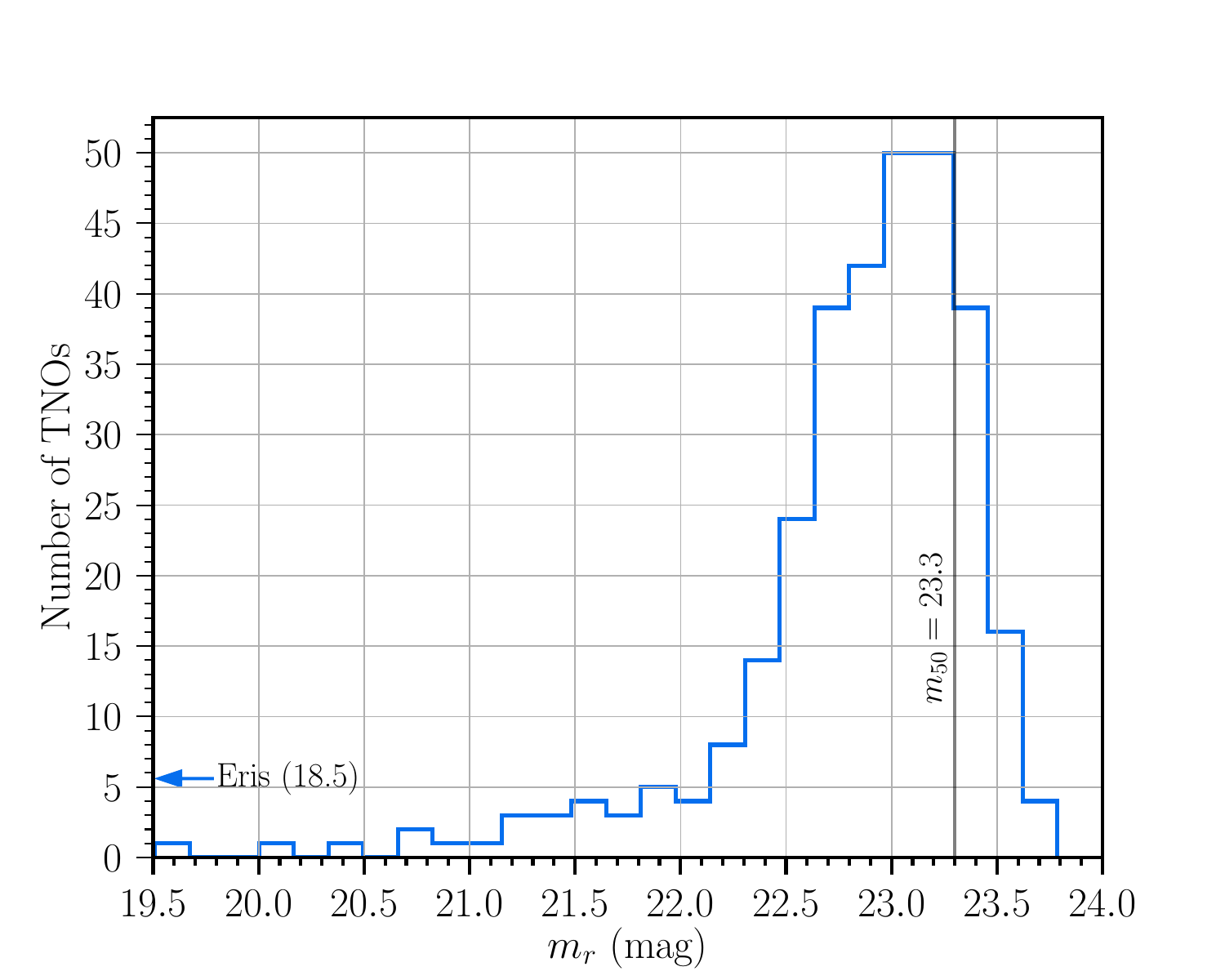}
	\includegraphics[width=0.49\textwidth]{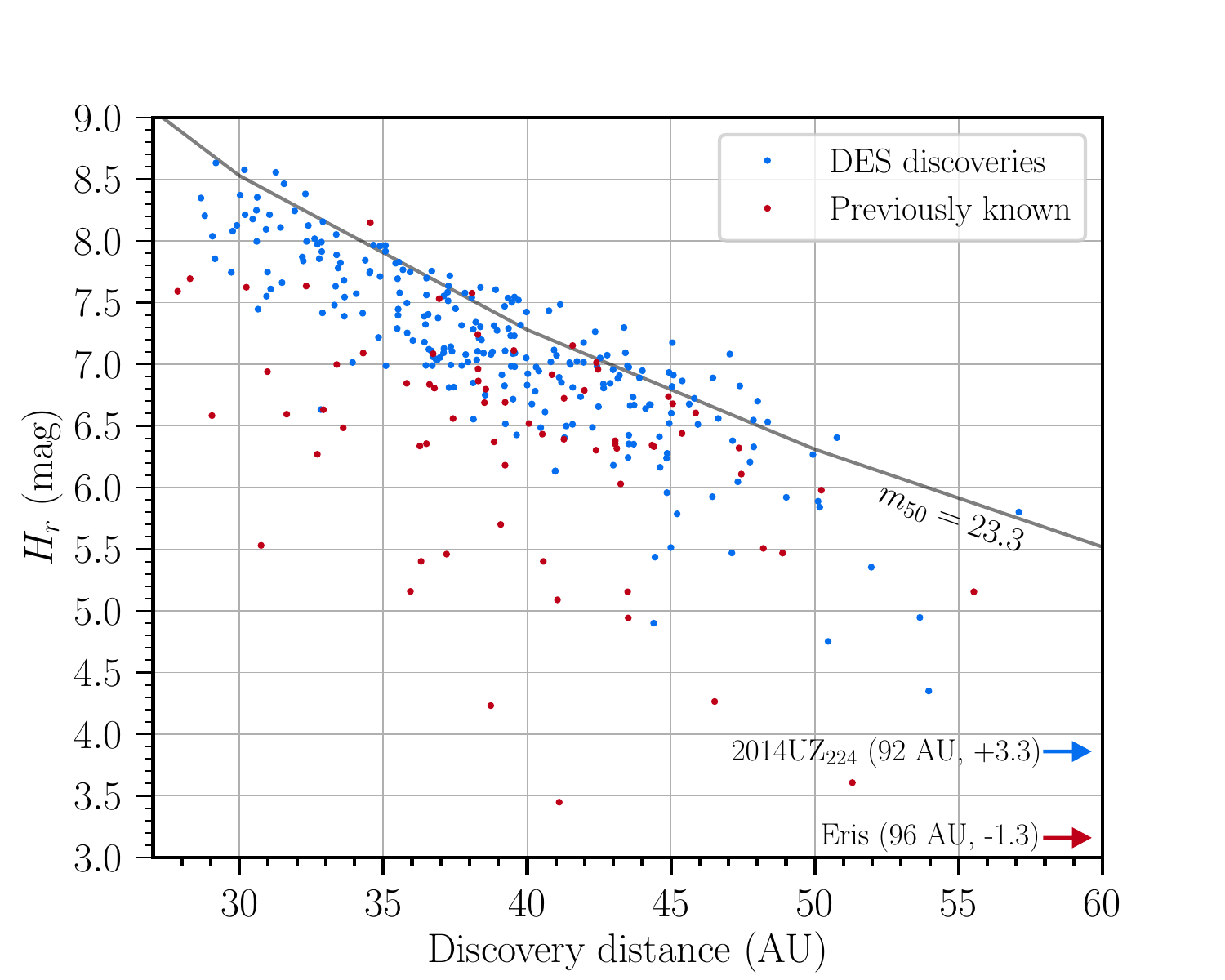}
	\caption{\emph{Left:} Histogram of apparent $r$ band magnitudes for all recovered objects, with the black line indicating the magnitude of 50\% completeness of the survey at $m_{50} = 23.3$. \emph{Right:} Discovery distance vs.\ absolute $r$ magnitude for all objects, color coded to indicate whether the object was known before \des\ or not. Note that, in both figures, Eris is indicated to be outside the plot's range. 2014 UZ$_{224}$ is outside the range of the second figure as well. The magnitudes ($m_r$ and $H_r$) and distances for these objects are indicated in parenthesis.\label{im:absmag}}
\end{figure}

\begin{figure}[tb]
	\centering
	\includegraphics[width=0.8\textwidth]{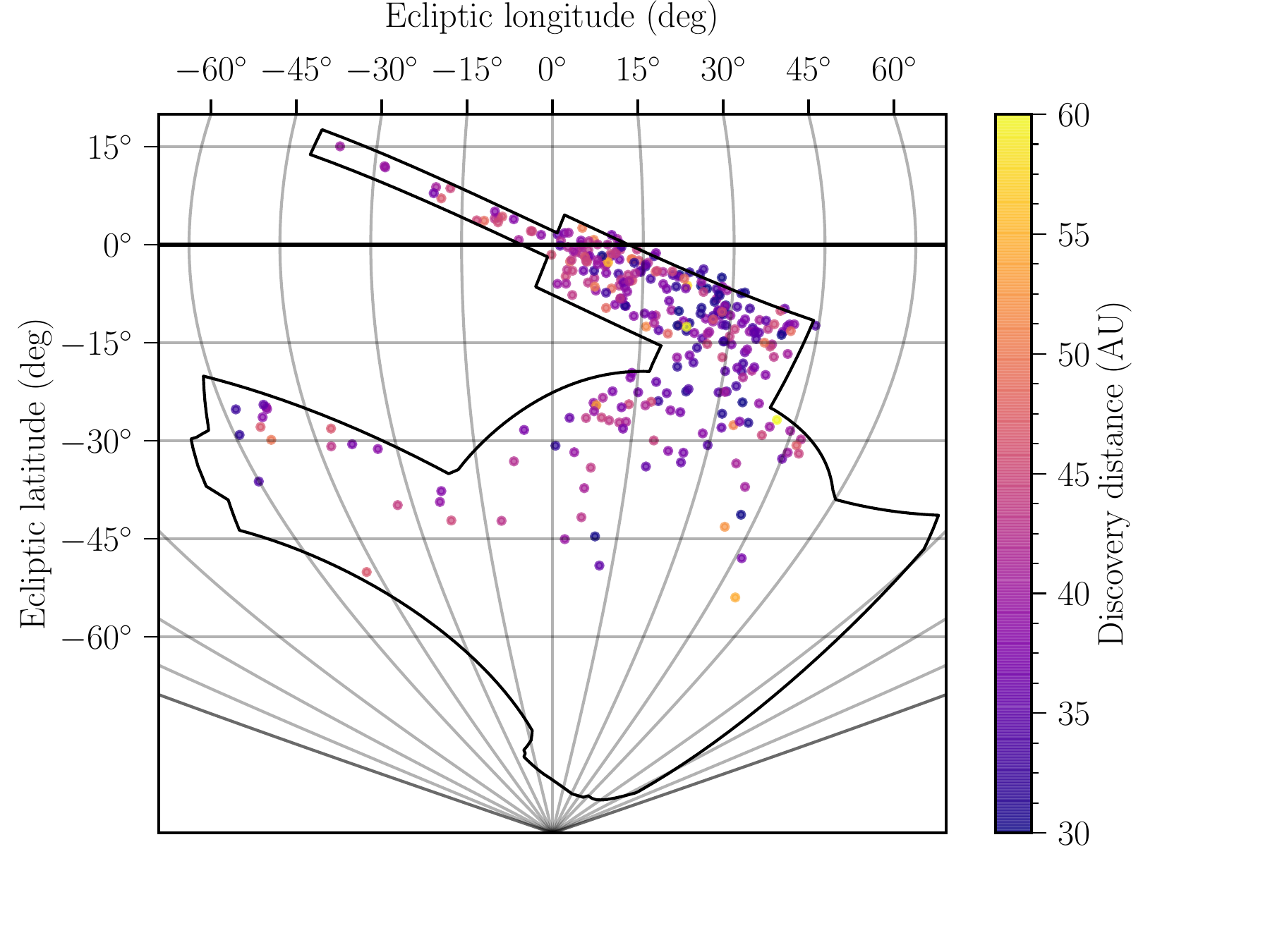}
	\caption{Location of the TNOs reported here at the epoch
          $\mathrm{JD} = 2457388.4$ (January $1^\mathrm{st}$,
          2016). The color scale represents the discovery distance
          (note that two of the objects were found at $d > 90 \,
          \mathrm{AU}$). The black line bounds the survey footprint.
          Note the paucity of TNO detections at ecliptic latitudes below $-50\degr.$
          \label{im:tno_map}}
\end{figure}

The semi-major axes, eccentricities and inclinations for all Y4 objects are plotted in Figure~\ref{im:aei}, their magnitudes in Figure~\ref{im:absmag}, and their locations for a fixed epoch in Figure~\ref{im:tno_map}.  We highlight the following properties of our sample:
\begin{itemize}
\item Of the 316 objects reported here, 139 are reported here for the first time, and 245 are \des\ discoveries. In Table \ref{tb:pop}, we further divide the objects by dynamical classes following \cite{Gladman2008}. The full methodology is presented in \cite{Khain2019}. The classification is made by integrating ten clones of each object for 10 Myr, and resonances of the form $p:q$, $p,q\in[1,26]$ are identified using an automated system.
\item We have 54 detached objects in our sample, one of the largest samples of this population to date, although some of the detached objects near high order resonances with Neptune might end up reclassified as resonant as the orbits are refined.
\item There are 87 confirmed resonant objects, with another 13 resonant candidates. Of the confirmed objects, 7 are Neptune Trojans (four of them new to this work), 30  plutinos and two in the 4:1 resonance. There is one 16:3 resonant candidate ($a \sim 91.1 \, \mathrm{AU}$).  
\item There are seven extreme TNOs ($a > 150 \, \mathrm{AU}$, $q > 30 \, \mathrm{AU}$), including one $a > 250 \, \mathrm{AU}$ object new to this work.
\item Since most of our area is far from the ecliptic, almost half of our sample consists of objects with $i > 20\degr$. 
\item The flattened distribution of the Kuiper belt is readily apparent, and indeed we have discovered no TNOs more than 55\arcdeg\ from the ecliptic plane despite substantial sensitivity outside this range (compare Figure \ref{im:tno_map} to Figure~\ref{im:recovery}).
\item The truncation of the classical Kuiper Belt \citep{Allen2001} at the 2:1 resonance is also apparent, demonstrated by the small absence of low-eccentricity objects past 48 AU (upper panel of Figure~\ref{im:aei}).
\item The sample contains several extreme TNOs and several high-inclination ($i > 40 \degr$) objects that are difficult to produce in basic formation scenarios, and should be highly constraining for alternative dynamical scenarios.
\end{itemize}

\begin{deluxetable}{cccc}
	\tablecaption{\des\ trans-Neptunian objects. The description of each column is given here. The table will be provided in a machine-readable format. Each uncertainty is the 1-$\sigma$ uncertainty marginalized over all other orbital parameters. All of the elements reported are barycentric and refer to epoch 2016.0. \textbf{Table provided in FITS as an ancillary file or in the journal version of the paper.}\label{tb:catalog}}
	\tablehead{\colhead{Column name} & \colhead{Symbol} & \colhead{Unit} & \colhead{Description}}
	\startdata
	\texttt{MPC} & & & Minor Planet Center object designation \\
	\texttt{DES} & &  & DES object designation \\ 
	\texttt{a} & $a$  &  AU & Semi-major axis of the best-fit orbit \\
	\texttt{sigma\_a} & $\sigma_a$ & AU & Uncertainty in $a$ \\
	\texttt{e} & $e$ & & Eccentricity \\
	\texttt{sigma\_e} & $\sigma_e$ & & Uncertainty in $e$ \\
	\texttt{i} & $i$ & $\deg$ & Inclination \\
	\texttt{sigma\_i} & $\sigma_i$ & $\deg$ & Uncertainty in $i$ \\ 
	\texttt{aop} & $\omega$ & $\deg$ & Argument of perihelion \\
	\texttt{sigma\_aop} & $\sigma_\omega$ & $\deg$ & Uncertainty in $\omega$ \\
	\texttt{lan} & $\Omega$ & $\deg$ & Longitude of ascending node \\
	\texttt{sigma\_lan} & $\sigma_\Omega$ & $\deg$ & Uncertainty in $\Omega$ \\
	\texttt{T\_p} & $T_p$ & UTC Modified Julian date & Time of perihelion passage  \\ 
	\texttt{sigma\_T} & $\sigma_T$ & days & Uncertainty in $T_p$\\
	\texttt{q} & $q$ & AU & Perihelion distance \\
	\texttt{sigma\_q} & $\sigma_q$ & AU & Uncertainty in $q$ \\
	\texttt{d} & $d$ & AU & Discovery distance (geocentric) \\ 
	\texttt{sigma\_d} & $\sigma_d$ & AU & Uncertainty in $d$ \\
	\texttt{m\_r} & $m_r$ & mag & Mean $r$ band magnitude \\
	\texttt{sigma\_m} & $\sigma_m$ & mag & Uncertainty in $m_r$ \\
	\texttt{H\_r} & $H_r$ & mag & Absolute magnitude in band $r$ \\
	\texttt{sigma\_H} & $\sigma_H$ & mag & Uncertainty in $H_r$ \\
	\texttt{NUNIQUE} & $\mathtt{NUNIQUE}$ & & Number of unique nights of detections \\
	\texttt{NDETECT} & $\mathtt{NDETECT}$ & & Number of detections \\ 
	\texttt{CHI2} & $\chi^2$ & & $\chi^2$ of the orbit fit ($\nu = 2\times \mathtt{NDETECT} - 6$) \\ 
	\texttt{x}, \texttt{y}, \texttt{z} (3 columns) & $x,y,z$ & AU & ICRS-oriented positions\\
	\texttt{v\_x}, \texttt{v\_y}, \texttt{v\_z} (3 columns) & $v_x, v_y, v_z$ & $\text{AU}/\text{year}$ & ICRS-oriented velocities \\
 	\texttt{Sigma\_mu\_nu} (21 columns) & $\Sigma_{\mu,\nu}$  & $(\text{AU},\text{AU}/\text{yr})^2$ & $\mu,\nu$ element of the state vector covariance matrix.\\
	\texttt{Class} & Class & & Dynamical classification  \\
	\enddata 
\end{deluxetable}

\begin{deluxetable}{cccc}
	\tablecaption{Known trans-Neptunian objects found in other searches of \des\ data. The details of this search are explained elsewhere \citep{Khain2019}. \textbf{Table provided in FITS as an ancillary file or in the journal version of the paper.}\label{tb:umich}}
	\tablehead{\colhead{Column name} & \colhead{Unit} & \colhead{Description}}
	\startdata
	\texttt{MPC} & & & Minor Planet Center object designation \\
	\texttt{a} & $a$ & AU & Semi-major axis of the best-fit orbit \\
	\texttt{sigma\_a} & $\sigma_a$ & AU & Uncertainty in $a$ \\
	\texttt{e} & $e$ & & Eccentricity \\
	\texttt{sigma\_e} & $\sigma_e$ & & Uncertainty in $e$ \\
	\texttt{i} & $i$ & $\deg$ & Inclination \\
	\texttt{sigma\_i} & $\sigma_i$ & $\deg$ & Uncertainty in $i$ \\ 
	\texttt{aop} & $\omega$ & $\deg$ & Argument of perihelion \\
	\texttt{sigma\_aop} & $\sigma_\omega$ & $\deg$ & Uncertainty in $\omega$ \\
	\texttt{lan} & $\Omega$ & $\deg$ & Longitude of ascending node \\
	\texttt{sigma\_lan} & $\sigma_\Omega$ & $\deg$ & Uncertainty in $\Omega$ \\
	\texttt{T\_p} & $T_p$ & UTC Modified Julian date & Time of perihelion passage \\ 
	\texttt{sigma\_T} & $\sigma_T$ & days & Uncertainty in $T_p$\\
	\texttt{Reason} &  & & Reason the object was missed$^1$ \\
	\enddata 
	\tablenotetext{1}{d: deep fields, m: missing from transient catalog, l: failed linkage, c: geocentric distance closer than $30 \, \mathrm{AU}$, s: short arc (i.e. $\mathtt{ARCCUT} < 6$~months)} 
\end{deluxetable}
\begin{deluxetable}{cc}
	\tablecaption{Dynamical classification of the 316 objects following \cite{Gladman2008} \citep[see][for details]{Khain2019}. The resonant objects are presented in order of increasing semi-major axis, with the approximate value presented in parenthesis. \label{tb:pop}}
	\tablehead{\colhead{Dynamical class} & \colhead{Number of Objects} }
	\startdata
	Classical belt & 134 \\
	Scattering & 28 \\
	Detached & 54 \\
	\hline \hline
	Mean-motion resonators with Neptune & Number of objects \\ \hline
	1:1 (30.1 AU) & 7 \\
	5:4 (34.9 AU)  & 1 \\
	4:3 (36.3 AU) & 7 \\
	3:2 (39.4 AU) & 30 + 4 candidates \\
	5:3 (42.3 AU) & 6 \\
	12:7 (43.1 AU) & 1 candidate \\
	7:4 (43.7 AU) & 12 \\
	2:1 (47.7 AU) & 5 + 2 candidates \\
	21:10 (49.3 AU) & 1 \\
	13:6 (50.4 AU) & 1 \\
	9:4 (51.7 AU) & 1 candidate \\
	7:3 (52.9 AU) & 1 + 2 candidates \\
	5:2 (55.4 AU) & 8 + 1 candidate \\
	3:1 (62.6 AU) & 1 \\
	16:5 (65.4 AU) & 1 \\
	10:3 (67.1 AU) & 1 \\
	7:2 (69.4 AU) & 3 + 1 candidate\\
	4:1 (75.8 AU) & 2 \\
	16:3 (91.9 AU) & 1 candidate\\
	\hline \hline 
	Total & 316\\
	\enddata
\end{deluxetable}

Detailed characterization of the TNO populations will be presented in future publications.

\section{Summary and prospects}
\label{sec:conclusions}
We report 316 trans-Neptunian objects found in the first four years of data from the $5000 \, \deg^2$ Dark Energy Survey. The astrometry for this sample was calibrated to Gaia Data Release 1. This search is complete to magnitude $m_r \approx 23.3$.   \des\ is a temporally sparse survey, requiring us to develop new methods to identify moving objects and link them into orbits.  A technique applied here to TNOs for the first time (to our knowledge) is to confirm orbital discoveries using the ``sub-threshold significance'' statistic, whereby we stack along the orbit using only exposures that are statistically independent of the ones used to discover the object.  This provides a very clear distinction between real sources and spurious linkages.  The need for this arises from the fact that, in a survey as large as \des, distinct asteroid detections or defects can align so as to mimic a true TNO on as many as 7 distinct nights spread over multiple years.  

The large contiguous field and homogeneous coverage of \des\ is shown to yield a selection function that is nearly independent of orbital elements, as long as the orbit and source magnitude yield detections on at least 6 \des\ exposures spanning multiple seasons.  This will make it relatively straightforward to compare the \des\ Y4 catalog to candidate models of TNO populations.  Each detected object has already been observed in 5 filters with multiple years' arc, and the survey spans a large range of ecliptic latitude, making it valuable for  comparisons to theory.  We plan, however, to defer the most detailed comparisons until we apply our methods to the full \des\ dataset.

We expect many improvements for the final analysis of \des\ data from the full six years of the survey. The \texttt{SExtractor} \citep{Bertin1996} detection filter has been changed to better approximate the PSF for \des\ images, and the detection threshold has been lowered.  These should yield 0.3 mag fainter $m_{50}$ with only modest increase in the size of transient catalog. We also will have full 10-tiling coadds, leading to more efficient rejection of stationary objects. With the six years of data, we expect that most orbit arcs will all be at least 3 years long, and our photometry will be improved from multiple epochs of data. In addition, our astrometry will be calibrated to Gaia Data Release 2 \citep{GaiaDR2}.

\acknowledgments

\emph{Software}: The software developed in this work will be made public shortly after the publication. This work made use of the following public codes: \textsc{Numpy} \citep{Numpy}, \textsc{SciPy} \citep{SciPy}, \textsc{Astropy} \citep{Astropy2013,Astropy2018}, \textsc{Matplotlib} \citep{Matplotlib}, \textsc{IPython} \citep{iPython}, \textsc{TreeCorr} \citep{Jarvis2003}, \textsc{easyaccess} \citep{Easyaccess}, \textsc{WCSFit} and \textsc{pixmappy} \citep{Bernstein2017astro}, \textsc{SExtractor} \citep{Bertin1996}, \textsc{CFITSIO} \citep{Cfitsio}, \textsc{Eigen} \citep{eigenweb}, \textsc{CSPICE} \citep{SPICE,ACTON20189}, \textsc{SMP} \citep{Brout2019}

University of Pennsylvania authors have been supported in this work by grants AST-1515804 and AST-1615555 from the National Science Foundation, and grant DE-SC0007901 from the Department of Energy. Work at University of Michigan is supported by the National Aeronautics and Space Administration under Grant No. NNX17AF21G issued through the SSO Planetary Astronomy Program and  NSF Graduate Research Fellowship Grant No. DGE 1256260.

Funding for the DES Projects has been provided by the U.S. Department of Energy, the U.S. National Science Foundation, the Ministry of Science and Education of Spain, 
the Science and Technology Facilities Council of the United Kingdom, the Higher Education Funding Council for England, the National Center for Supercomputing 
Applications at the University of Illinois at Urbana-Champaign, the Kavli Institute of Cosmological Physics at the University of Chicago, 
the Center for Cosmology and Astro-Particle Physics at the Ohio State University,
the Mitchell Institute for Fundamental Physics and Astronomy at Texas A\&M University, Financiadora de Estudos e Projetos, 
Funda{\c c}{\~a}o Carlos Chagas Filho de Amparo {\`a} Pesquisa do Estado do Rio de Janeiro, Conselho Nacional de Desenvolvimento Cient{\'i}fico e Tecnol{\'o}gico and 
the Minist{\'e}rio da Ci{\^e}ncia, Tecnologia e Inova{\c c}{\~a}o, the Deutsche Forschungsgemeinschaft and the Collaborating Institutions in the Dark Energy Survey. 

The Collaborating Institutions are Argonne National Laboratory, the University of California at Santa Cruz, the University of Cambridge, Centro de Investigaciones Energ{\'e}ticas, 
Medioambientales y Tecnol{\'o}gicas-Madrid, the University of Chicago, University College London, the DES-Brazil Consortium, the University of Edinburgh, 
the Eidgen{\"o}ssische Technische Hochschule (ETH) Z{\"u}rich, 
Fermi National Accelerator Laboratory, the University of Illinois at Urbana-Champaign, the Institut de Ci{\`e}ncies de l'Espai (IEEC/CSIC), 
the Institut de F{\'i}sica d'Altes Energies, Lawrence Berkeley National Laboratory, the Ludwig-Maximilians Universit{\"a}t M{\"u}nchen and the associated Excellence Cluster Universe, 
the University of Michigan, the National Optical-Infrared Astronomy Observatory, the University of Nottingham, The Ohio State University, the University of Pennsylvania, the University of Portsmouth, 
SLAC National Accelerator Laboratory, Stanford University, the University of Sussex, Texas A\&M University, and the OzDES Membership Consortium.

Based in part on observations at Cerro Tololo Inter-American Observatory, National Optical-Infrared Astronomy Observatory, which is operated by the Association of 
Universities for Research in Astronomy (AURA) under a cooperative agreement with the National Science Foundation.

The DES data management system is supported by the National Science Foundation under Grant Numbers AST-1138766 and AST-1536171.
The DES participants from Spanish institutions are partially supported by MINECO under grants AYA2015-71825, ESP2015-66861, FPA2015-68048, SEV-2016-0588, SEV-2016-0597, and MDM-2015-0509, 
some of which include ERDF funds from the European Union. IFAE is partially funded by the CERCA program of the Generalitat de Catalunya.
Research leading to these results has received funding from the European Research
Council under the European Union's Seventh Framework Program (FP7/2007-2013) including ERC grant agreements 240672, 291329, and 306478.
We acknowledge support from the Australian Research Council Centre of Excellence for All-sky Astrophysics (CAASTRO), through project number CE110001020, and the Brazilian Instituto Nacional de Ci\^encia
e Tecnologia (INCT) e-Universe (CNPq grant 465376/2014-2).

This manuscript has been authored by Fermi Research Alliance, LLC under Contract No. DE-AC02-07CH11359 with the U.S. Department of Energy, Office of Science, Office of High Energy Physics. The United States Government retains and the publisher, by accepting the article for publication, acknowledges that the United States Government retains a non-exclusive, paid-up, irrevocable, world-wide license to publish or reproduce the published form of this manuscript, or allow others to do so, for United States Government purposes.

This research used resources of the National Energy Research Scientific Computing Center (NERSC), a U.S. Department of Energy Office of Science User Facility operated under Contract No. DE-AC02-05CH11231.

This work has made use of data from the European Space Agency (ESA)
mission {\it Gaia} (\url{https://www.cosmos.esa.int/gaia}), processed by
the {\it Gaia} Data Processing and Analysis Consortium (DPAC,
\url{https://www.cosmos.esa.int/web/gaia/dpac/consortium}). Funding
for the DPAC has been provided by national institutions, in particular
the institutions participating in the {\it Gaia} Multilateral Agreement.

\bibliography{references}
\bibliographystyle{aasjournal}

\end{document}